\PassOptionsToPackage{table}{xcolor}
\documentclass[manuscript,screen,nonacm,oneside]{acmart}
\usepackage[tableposition=below]{caption}
\usepackage{array}
\usepackage{multirow}
\usepackage{calc}
\usepackage{soul}
\usepackage{colortbl}
\usepackage{longtable}
\usepackage{listings}
\definecolor{codegray}{gray}{0.96}
\lstdefinelanguage{json}{
  basicstyle=\small\ttfamily,
  string=[s]{"}{"},
  stringstyle=\color[HTML]{770a2a},
  comment=[l]{//},
  commentstyle=\color{gray},
  literate=
    *{:}{{{\color[HTML]{1f729e}:}}}{1}
     {,}{{{\color[HTML]{1f729e},}}}{1}
}
\definecolor{color2}{HTML}{ad0015}
\definecolor{rowgraydark}{gray}{0.85}
\definecolor{color4}{HTML}{f2a7af}
\definecolor{color5}{HTML}{cddfe5}
\definecolor{rowgray}{HTML}{F2F4F7}
\AtBeginDocument{%
  }

\setcopyright{none}
\hypersetup{    
      colorlinks=true,
      linkcolor=blue,
      citecolor=blue,
      urlcolor=blue
  }
\begin{document}

\title[How are AI agents used? Evidence from 177,000 MCP ...]{How are AI agents used? Evidence from 177,000 MCP tools}

\author{Merlin Stein}
\affiliation{%
  \institution{UK AI Security Institute, University of Oxford}
  \country{UK}}
\email{merlin.stein@dsit.gov.uk}

\renewcommand{\shortauthors}{Author}

\begin{abstract}
Today's AI agents are built on large language models (LLMs) equipped with tools to access and modify external environments, such as corporate file systems, API-accessible platforms and websites. AI agents offer the promise of automating computer-based tasks across the economy. However, developers, researchers and governments lack an understanding of how AI agents are currently being used, and for what kinds of (consequential) tasks. To address this gap, we evaluated 177,436 agent tools created from 11/2024 to 02/2026 by monitoring public Model Context Protocol (MCP) server repositories, the current predominant standard for agent tools. We categorise tools according to their direct impact: \emph{perception tools} to access and read data, \emph{reasoning tools} to analyse data or concepts, and \emph{action tools} to directly modify external environments, like file editing, sending emails or steering drones in the physical world. We use O*NET mapping to identify each tool's task domain and consequentiality. Software development accounts for 67\% of all agent tools, and 90\% of MCP server downloads. Notably, the share of 'action' tools rose from 27\% to 65\% of total usage over the 16-month period sampled. While most action tools support medium-stakes tasks like editing files, there are action tools for higher-stakes tasks like financial transactions. Using agentic financial transactions as an example, we demonstrate how governments and regulators can use this monitoring method to extend oversight beyond model outputs to the tool layer to monitor risks of agent deployment.
\end{abstract}



\maketitle

\section{Introduction}
\label{sec:introduction}

The field of artificial intelligence is moving from systems that generate content towards systems that can perceive their environment and act upon it \citep{russell2020artificial} in an increasingly autonomous fashion (AI agents).
Today's AI agents are systems built on Large Language Models (LLMs), augmented with memory components and provided with access to external tools.
Tools are functions that the agent can call to access, analyse or modify external environments. Unlike a standard LLM that can draft correspondence, list products, or recommend stocks, an agent can use tools to autonomously send an email (e.g. using Google MCP), search an online marketplace (e.g. using Amazon MCP), or execute financial transactions (e.g. using Coinbase MCP), often with relatively little human oversight. This capacity for autonomous environmental modification means that AI agents create new risks beyond those of standard LLMs \citep{cihon2025measuring}. These include security vulnerabilities in agents or their tools that compromise functionality, the misuse of agents to orchestrate cyberattack campaigns, and agents that modify environments in a manner divergent from what a user intended. Agent tools also raise novel structural risks, including the risk of cascading effects when many agents complete tasks simultaneously or in a coordinated manner \citep{ALDASORO2025101472,danielsson2025artificialintelligencefinancialcrises}. However, to date, we have limited insight into what agent tools are being developed, what tasks they enable, and which agent tools are widely used.

Most studies that monitor how AI systems are being used focus on usage of LLMs \citep{chatterji2025people,handa2025which}. User surveys often fail to distinguish between LLM usage and AI agent usage \citep{hai2025aiindex,googleRealworldCases,pwc2024agents}. More recently, there have been attempts to track agent usage through platform data. \citep{yang2025adoption} find productivity and learning dominate AI agent usage. \citep{aubakirova2025state} find software development is the focus for the majority of agent queries; \citep{casper2025agent}'s mapping of 67 agents finds the same. \citep{pan2025measuring} survey 306 implementers to learn that current agent deployments focus on tasks with few steps. However, we lack large-scale evidence about how AI agents are currently being used across platforms via tool calls, and how agents are interacting with external environments through tools. 

To address this, this paper classifies AI agent tools and measures their popularity by tracking Model Context Protocol servers (MCPs). An MCP server is a lightweight program that exposes capabilities and external environments to AI agents via a standardised protocol, like data sources, APIs, or a browser. Each MCP server packages one or more tools ('MCP tools'). Most large agent developers, like OpenAI, Anthropic and Google \citep{casper2025agent}, provide MCP integrations. MCP is currently the dominant open protocol for agent tools -- all agent-related repositories in GitHub's top 10 new repositories of H1/2025 build MCP infrastructure or integrate MCP \citep{clickhouseClickHouseQuery,hackeroneWhatModel}. MCP tools are published by both agent developers, like Anthropic; digital service providers, like Asana; established digitised businesses, like Griffin Bank; and open-source developers. Monitoring MCP servers (where tools are exposed to agents) gives us insight into which tools are published, downloaded, and thus which tasks agents are performing in the wild.

Prior work provides descriptive snapshots of MCP servers \citep{lin2025large,ray2025survey,hou2025modelcontextprotocolmcp}, finding most are developed with Python. Instead, our work uses automated classification of the tools on MCP servers to provide time-varying information about how AI agents are being used. We analyse tool capabilities based on their direct impact (perception, reasoning, or action) and whether tools provide access to narrow, constrained environments (like executing an action with a specific API) or general, unconstrained environments (like manipulating a web browser). We map these tools to work-related task domains using the O*NET occupational taxonomy to assess the consequentiality of the tasks. This paper contributes:

\begin{enumerate}
\def\labelenumi{\arabic{enumi}.}
\item
 \textbf{An overview of the agent tool ecosystem.} We curate an agent tool dataset of 177,436 public MCP tools (available upon request), the largest dataset of AI Agent tools existing to date, sourced from MCP servers on GitHub and Smithery. We categorise which tasks agent tools help complete, and the stakes of these tasks.
\item
 \textbf{Time trends of agent tool usage.} We track over time whether agent tools are being created to enable perception, reasoning, or action. We augment our dataset of public tools with data on the number of downloads from the Node Package Manager (NPM) and Python Package Index (PyPI), covering 3,854 MCP servers (42,498 tools), to approximate usage trends. 
\item
 \textbf{A method for early monitoring of wide or high-stakes AI agent deployment, to anticipate risks}. We show how monitoring public agent tools helps anticipate and monitor large-scale agent deployment, and prepare for potential opportunities and risks \citep{stein2024role,bernardi2024societal}. We offer the example of tracking agentic transaction tools, that was part of a trial monitoring project with the UK government's financial authorities.
\end{enumerate}

We report several descriptive results:

\begin{enumerate}
\def\labelenumi{\arabic{enumi}.}
\item
 \textbf{Software development and other IT tools currently dominate}. We find that 67\% of published tools (90\% of usage) are software-related, with tools for financial and administrative tasks also proving popular (Table~\ref{tab:domains}).
\item
 \textbf{AI agent actions seem to be concentrated in the United States}. Approximately 50\% of all tool usage is in the US, followed by Western Europe (\textasciitilde20\%) and China (\textasciitilde5\%) (Figure~\ref{fig:geographic-distribution}). We track the IP addresses of agent tool downloads from the Western-focused PyPI, but do not observe whether downloaded tools are actually called.
\item
 \textbf{Over time, there has been a shift towards tools that let agents make direct modifications, in unconstrained, general environments} (Figure~\ref{fig:action-tools-trend}). Uses of action tools grew from 27\% (11/2024) to 65\% (02/2026) of total tool uses (action, perception, reasoning tools). This is driven by an increase in general-purpose tools that permit access to unconstrained environments, enabling an agent to use a computer or browser (Figure~\ref{fig:generality-tools}).
\item
 \textbf{Tools that permit `actions' are associated mainly with medium-stakes occupations}. However, financial transactions are one area with fast growth of potentially high-stakes agent actions (Figure~\ref{fig:stakes-occupations}).
\item
 \textbf{A significant and growing share of AI agent tools are created with the help of AI agents}. We detect AI assistance in 28\% of MCP servers (36\% of tools). The share of new MCP servers created with AI assistance rose from 6\% (01/2025) to 62\% (02/2026), dominated by Claude Code (69\% of AI-coauthored servers) (Figure~\ref{fig:ai-created-results}).
\end{enumerate}

Section~\ref{sec:background} establishes why monitoring agent tools can anticipate risks of AI agents. Section~\ref{sec:data} describes our panel data on tools and their use. Sections~\ref{sec:methodology} and~\ref{sec:results} present methods and findings, Section~\ref{sec:discussion} discusses risk implications and Section~\ref{sec:conclusions} concludes.

\section{Background}
\label{sec:background}
\subsection{Characterising the action space of AI agents}
AI agents mark a paradigm shift in AI, enabling faster, cheaper, and wider-ranging computer-based actions. For example, while LLMs can provide information on different investment strategies, AI agents can execute trades directly according to that strategy using a series of autonomous computer based commands and external tools. This means that AI agents can create and execute novel, adaptive trading strategies faster and at a lower cost than human operators, and more adaptively than traditional narrow trade-execution software \citep{li2025hedgeagents}. 

This capacity of AI agents to directly fulfil a wide variety of computer-based tasks depends on their autonomy, goal complexity and action space \citep{kasirzadeh2025characterizing, chan2023harms, imda}. 

\textbf{Autonomy} is the ability of an AI agent to fulfil tasks without external direction or controls, e.g. follow-up prompts by humans or other AI systems \citep{cihon2025measuring}. An AI agent's autonomy is determined not by what tools it has access to, but how the system is configured. For example, when given a command, personal assistants like Alexa and Siri are configured to seek a human user's input for most steps. By contrast, coding agents like Claude Code can be configured to complete multiple steps and actions from a single prompt, giving it a higher degree of autonomy (e.g. default permissions in 'settings.json', 'dangerously-skip-permissions' mode).

\textbf{Goal complexity} is the ability of an AI agent to pursue high-level objectives through goal decomposition and adaptive planning over extended time periods \citep{chan2023harms, kasirzadeh2025characterizing}. An agent's goal complexity depends on the size of its context window, the ability of its memory system to bridge across context windows, and the capability of its underlying large language model to choose suitable steps to solve a problem - not on its tools. For example, customer service agents running on older LLMs like GPT-4 typically handle simple requests without retaining the context of previous interactions or steps the system took. An AI scientist agent running on Claude Opus 4.6 or GPT-5.2, by contrast, effectively uses its layered memory systems to maintain context and progress steadily across extended research workflows \citep{lu2024aiscientist}.

The \textbf{action space} of an AI agent describes the set of actions it can take in the world. Tools define an AI agent's action space. For example, ChatGPT can only access a user's Google Drive, if the user enables a dedicated Google Drive tool, a browser tool (and provides Google login credentials) or a code execution tool (and provides API tokens).

Other researchers have reviewed how the autonomy \citep{cihon2024chilling} and goal complexity \citep{chan2023harms} of AI agents may influence risks. In this paper, we focus on the action space \citep{imda}. 

\subsection{The action space shapes the risks of AI agents}

While safety concerns regarding LLMs have historically focussed on information hazards like the generation of harmful or inaccurate content \citep{bommasani2021opportunities}, AI agents amplify risks of misuse, misalignment, mistakes and structural harms as their autonomy, goal complexity and action space expands \citep{shah2025approach, bengio2025superintelligent}. The following provides an overview of how understanding the action space of agents can shed light on risks in each of these categories.

\textbf{Misuse risks} occur when malicious actors exploit AI agents for harmful purposes. Cyber criminals have manipulated AI agents to leak sensitive information or transfer cryptocurrency worth hundreds of thousands of dollars \citep{reddy2025echoleakrealworldzeroclickprompt, decryptAiXBTToken, akinciborgHackingCrypto, deng2025ai}. The attack surface of an AI agent is defined by its action space: First, an AI agent is more likely to encounter manipulative instructions (`prompt injections') when it accesses the open web or uses other general-purpose tools. Second, an AI agent can only act upon malicious instructions, when it has tools to act, e.g., to send emails or cryptocurrency \citep{mo2025attractive, hou2025modelcontextprotocolmcp, gan2024navigating}. 

Agent tools for some domains lower the cost of crime. General-purpose, dual-use tools for executing code have been exploited for cyber espionage in 2025 \citep{anthropicDisruptingFirst}. Narrow-purpose tools, such as password brute-forcing tools, could further lower barriers if widely adopted \citep{ncscImpactCyber}. Monitoring whether such tools see scaled adoption offers an early signal of changing criminal capabilities.

\textbf{Mistake and misalignment risks} occur when AI agents take erroneous or unintended actions. In simulations, misaligned agents have blackmailed their users \citep{anthropicAgenticMisalignment}. Mistakes of production agents have deleted live databases and exposed hundreds of thousands of patient records \citep{fortuneAIpoweredCoding, abaSleepwalkingInto}. These incidents happened because AI agents had tools to do irreversible actions: they could send emails, execute code and modify databases. Agents with high-stakes tools for actions, like cryptocurrency transfer tools, can cause immediate, irreversible financial damage \citep{hammond2025multiagentrisksadvancedai}. Misaligned or erroneous agents limited to reasoning tools can deceive users, but not act themselves. The LLM underlying an agent determines an agent's tendency for misaligned behavior - whether this tendency turns into harmful actions depends on the tools available to an agent.  

Large-scale usage of such high-stakes tools for modifications amplifies risk. Many agents run on the same underlying LLMs; an alignment failure in a widely-deployed model propagates to every agent built on it \citep{vipra2023market}. If those agents have tools to act in high-stakes context and operate worldwide, a single problematic update could trigger correlated failures across financial systems or critical infrastructure \citep{tomasev2025virtual}. Some regulators already monitor the degree of dependency of critical entities on few large language models \citep{bankofenglandBankEnglands}. Monitoring which tools are available to AI agents and whether they are used by critical entities would complement that approach to ensure human oversight over consequential agent actions. 

\textbf{Structural risks} may arise from large-scale deployment of AI agents as imperfect substitutes for human operators \citep{kulveit2025gradual}. When AI agent actions replace human actions, these actions are often more correlated due to similar training, faster and, inherently, excluding human participation \citep{laurito2025aiai}. For example, AI agents equipped with tools to edit encyclopedia or other web pages at scale, might make the content that humans read online less diverse \citep{battista2024political, qiu2025lockin}. However, structural risks of AI agents are mostly theoretical and have not yet materialised.

Agent actions in unconstrained settings could crowd out human operators \citep{ibrahim2025measuring}. Cascades of agent actions could destabilise critical sectors \citep{ibrahim2025measuring,deng2025ai}. For instance, agents with general-purpose tools like browser\_click or phone\_call could flood government websites or emergency lines designed for human use \citep{kulveit2025gradual}.
Simultaneous use of high-stakes banking tools for automated withdrawals by financial agents could precipitate liquidity crises akin to `agent bank runs' \citep{ALDASORO2025101472,danielsson2025artificialintelligencefinancialcrises}. The scale of action tool usage across a wide range of tasks signals whether AI agents fulfil a similar range of economically valuable tasks compared to humans, and indicates potential impacts on labor markets \citep{motwani2024secret, chan2023harms, shao2025future}.
These structural risks may be compounded by recursive self-improvement: AI agents that create their own tools expand the action space without requiring human effort \citep{ishibashi-etal-2025-large, Barrett_2016, zhang2025darwingodelmachineopenended}. When AI coding agents build new tools for other AI agents, tool proliferation is no longer bottlenecked by human developers, and tool creation may scale beyond human oversight.

The review above implies that five attributes of agent tools expand the action space, and amplify risks: moving (1) from perception to action tools, (2) from constrained to unconstrained environments, (3) from low-stakes to high-stakes tool use, (4) from small-scale to wide-ranging tool use across task domains and geographies and (5) from human-authored to AI-authored. The following section characterises each attribute. 

\subsection{Characteristics of AI agents: Tools define an AI agent's action space}
\label{sec:characteristics}
We distinguish five attributes of tools that define the set of actions an agent can take in the world - its action space:

\begin{enumerate}
\def\labelenumi{\arabic{enumi}.}
\item
  The \textbf{direct impact} of an AI agent tool describes whether a tool permits perception, reasoning or action. Agents need \emph{perception tools} to access and read data, \emph{reasoning tools} to analyse data or concepts, and \emph{action tools} to directly modify external environments, like file editing, sending emails or steering drones in the physical world. Agents limited to sensor-style tools, but without tools to act, are limited in their direct impact.

\item 
  Tool \textbf{generality} describes whether the tool enables interaction with narrow, constrained or general, unconstrained environments. \emph{Narrow-purpose tools} enable agents to fulfil tasks in constrained environments, such as a tool designed exclusively for transferring a cryptocurrency or viewing data via a particular API. \emph{General-purpose tools} grant agents access to unconstrained environments, such as the ability to control a web browser or execute arbitrary code. 

\item 
  The \textbf{task domain} of an AI agent tool describes the typical kind of work the tool helps to fulfil. We classify domains using the O*NET framework of economic tasks and occupations, and distinguish lower-stakes from higher-stakes domains based on the consequentiality of occupations supported by a particular tool. A tool to submit feedback is less consequential than a tool to submit cryptocurrencies trades. 

\item 
  Tool usage \textbf{geography} describes the region where a tool is used. Some tools are built for and used in a single country, such as Cyprus-specific trading tools, while others are used worldwide, such as tools that manage AI-agent connections or enterprise workspaces.

\item
  Tool \textbf{AI co-authorship} describes whether a tool was created with the assistance of AI coding agents. Some tools are created fully by human developers, other tools are conceptualised by humans but created with the help of AI coding agents, and potentially, future tools may be conceptualised and created fully by AI assistants.
\end{enumerate}

\begin{table*}[htbp]
\centering
\caption{AI agent tool examples by generality and direct impact.}
\small
\resizebox{\textwidth}{!}{%
\begin{tabular}{>{\raggedright\arraybackslash}p{3cm}>{\raggedright\arraybackslash}p{4cm}>{\raggedright\arraybackslash}p{4cm}>{\raggedright\arraybackslash}p{4cm}}
\toprule
\textbf{} & \textbf{Perception} & \textbf{Reasoning} & \textbf{Action} \\
\midrule
\textbf{Narrow-purpose} & \cellcolor{color5}Search for tickets (Ticketmaster MCP) & \cellcolor{color5}Reason through biomedical research questions (BioMCP) & \cellcolor{color4}Send crypto (Coinbase MCP) \\
\textbf{General-purpose} & \cellcolor{color5}Search the internet (DuckDuckGo MCP) & \cellcolor{color5}Perform accessibility audits of any website (A11Y MCP) & \cellcolor{color2}\textcolor{white}{Use a computer via mouse clicks (Desktop Commander MCP)} \\
\bottomrule
\end{tabular}%
}
\begin{minipage}{\linewidth}
\scriptsize
\textit{Notes: }{Agent tools for action and unconstrained environments increase the action space of an AI agent the most (dark red). Narrow-purpose action tools increase an agent's action space somewhat (light red). Other tools less so (light grey).}
\end{minipage}
\label{tab:agency}
\end{table*}
\begin{table*}[htbp]
\centering
\caption{AI agent action tool examples by geography and consequentiality.}
\small
\resizebox{\textwidth}{!}{%
\begin{tabular}{>{\raggedright\arraybackslash}p{2.5cm}>{\raggedright\arraybackslash}p{4cm}>{\raggedright\arraybackslash}p{4cm}>{\raggedright\arraybackslash}p{4cm}}
\toprule
\textbf{} & \textbf{Low-stakes action} & \textbf{Medium-stakes action} & \textbf{High-stakes action} \\
\midrule
\textbf{One country} & \cellcolor{color5}Test software (mcp-server-spira, mainly US) & \cellcolor{color5}Build websites (django-mcp-server, mainly US) & \cellcolor{color5}Execute financial trades (metatrader-mcp-server, mainly Cyprus) \\
\textbf{One continent} & \cellcolor{color5}Bulk grade students, upload course materials (canvas-mcp, mainly N. America) & \cellcolor{color4}Create documentation (office-word-mcp-server, mainly Asia) & \cellcolor{color4}Configure delegation to AI agents (mcp-feedback-enhanced, mainly Asia) \\
\textbf{Worldwide} & \cellcolor{color5}Configure AI agent tool connections (llmling) & \cellcolor{color4}Manage enterprise software suite (mcp-server-odoo) & \cellcolor{color2}\textcolor{white}{Manage digital workspace permissions, send emails, ...  (google\_workspace\_mcp)} \\
\bottomrule
\end{tabular}%
}
\begin{minipage}{\linewidth}
\scriptsize
\textit{Notes: }{Tools used in high-stakes contexts, across geographies, indicate the largest action space (dark red), tools for medium-stakes actions use some geographies indicate medium action space (light red), tools used mainly in one country or able to only do low-stakes actions indicate smaller action space (light grey). Stakes are low ($<$50), medium (50-75) and high ($>$75) based on the O*NET 0-100 ranking of the impact of decisions of occupations which the tool supports. One country means $>$80\% of a tool's usage is in one country, worldwide means $<$70\% in one continent, one continent is the remainder. The most used MCP server is displayed for each cell.}
\end{minipage}
\label{tab:stakes}
\end{table*}
Tables \ref{tab:agency} and \ref{tab:stakes} illustrate how an agent's tools reveal the action space of an agent. A general-purpose action tool -- one that permits both access to unconstrained environments and direct action within them -- significantly expands an AI agent's action space (Table \ref{tab:agency}). With tools like 'computer\_mouse\_click' an AI agent is able to perform a wide range of computer-based tasks. 
Access to a variety of action tools for high-stakes tasks, to manage workspace permissions or bank accounts, expands an AI agent's action space into consequential territory. The scale of tool usage across regions proxies how widely agents are deployed (Table \ref{tab:stakes}).

Across risk types, tools that \textit{enable a large action space} for agents -- like general-purpose action tools -- amplify risks. AI agents \textit{operate in an extensive action space} when agents collectively use general-purpose action tools for high-stakes or wide-ranging tasks extensively \citep{mitchell2025fullyautonomousaiagents}. A large action space does not itself constitute harm -- but it allows for a wide range of actions. Some of these actions may be harmful individually; others are harmful in combination. Research can point to potential impacts and risks of AI agents by monitoring agent tools.

MCP tools published on developer platforms are \emph{early} indicators of agent tool trends and risks. For example, an unofficial Google Calendar MCP server was published on GitHub \citep{githubGitHubMarkelaugust74mcpgooglecalendar} on December 5, 2024, captured in our dataset. Anthropic \citep{Anthropic2025claude} and OpenAI \citep{openaiChatGPTRelease} added a pre-built Google Calendar tool to claude.ai and ChatGPT months later in April and August 2025. Large-scale downloads of MCP tools from developer platforms might foreshadow larger-scale usage in the wider population.

In five research questions (RQs), we investigate each of the attributes to empirically understand the action space of AI agents:
\begin{itemize}
    \item \textbf{RQ1:} How widely are AI agents used across domains?
    \item \textbf{RQ2:} How widely are AI agents used across geographies?
    \item \textbf{RQ3:} What fraction of AI agents are used for perception, reasoning, or action over time?
    \item \textbf{RQ4:} What fraction of AI agents are used to access narrow, constrained environments or general, unconstrained environments?
    \item \textbf{RQ5:} What fraction of AI agent tools are created with the support of AI agents?
\end{itemize}

\subsection{Current understanding of AI agent usage and action space}

Recent studies have tracked agent usage by domain, and partly by geography, but breakdowns of agent usage by generality and direct impact are scarce. Table \ref{tab:action_space_comparison} summarises the state of knowledge based on usage data studies of chatbots with perception tools and platforms offering agents with action tools. This includes analyses of Microsoft's Copilot chatbot equipped with search tools \citep{yang2025adoption}, Anthropic and OpenAI's chatbots with search and data analysis tools \citep{handa2025which} \citep{chatterji2025people}, OpenRouter's LLM inference and tool calling platform \citep{aubakirova2025state}, Anthropic's first-party LLM API marketed for agentic workflows \citep{handa2025which} and Perplexity's Comet agent with action tools for example to book flights \citep{yang2025adoption}.

Domains of usage are wide, while usage of agents with action tools is concentrated in computer-based task and occupations. \citep{aubakirova2025state} study usage of OpenRouter's model marketplace API, finding that most developers use agents for programming tasks (58\% of tokens). Data from the Claude API corroborates this focus on computer and mathematical tasks and occupations \citep{anthropic2026aeiv4}. Complementary large-scale evidence from Anthropic's analysis of Claude Code and public API usage shows that most agent activity remains concentrated in software engineering, with emerging experimentation in higher-stakes domains such as finance, healthcare, and cybersecurity \citep{anthropic2026agents}. \citep{casper2025agent} documented 67 deployed agentic systems, finding that 74.6\% specialise in software engineering or computer use, with 73.1\% developed by companies. Perplexity Comet browser agent is mostly used in knowledge-intensive occupations \citep{yang2025adoption}.

Practitioner and deployment studies have also examined agent development practices and deployment characteristics directly. \citep{pan2025measuring} surveyed 306 practitioners (including 86 with deployed agents) and conducted 20 in-depth interviews, finding agents are primarily deployed for tasks relating to technology (48\%), finance and banking (44\%), corporate services (42\%), and legal compliance (17\%) (N=69).

Agents are used worldwide, with adoption particularly high in countries with higher GDP per capita. Studies of chatbot and agent usage on claude.ai \citep{handa2025which}, ChatGPT \citep{chatterji2025people}, and Perplexity \citep{yang2025adoption} find a significant correlation between GDP per capita and usage. Half of OpenRouter usage \citep{aubakirova2025state} is concentrated in the US. However, these findings might not represent global usage patterns, as the existing literature focuses on US-based platforms, with limited evidence on usage distribution on Chinese or other platforms.

There is little evidence on the prominence of agents with action, reasoning or perception tooling. \citep{aubakirova2025state} study usage of OpenRouter's model marketplace API, finding that agent tool-calling rose to approximately 15\% of total tokens by late 2025 on their platform. \citep{kasirzadeh2025characterizing}, \citep{staufer20262025aiagentindex} and \citep{casper2025agent} characterise agentic systems by their ability to directly fulfil a wide variety of tasks along different dimensions. However, studies that distinguish the degree of direct impact of agent usage are missing.

Empirical evidence which environments agents access -- and whether those environments are constrained or unconstrained -- remains similarly limited. \citep{yang2025adoption} report the most frequently visited domains of the Comet web agent, such as google.com, offering a high-level view of browsing contexts without clarifying the scope of permissible actions. A survey of agent deployers \citep{pan2025measuring} highlights that 68\% deploy agents restrictively, typically allowing no more than ten steps before human intervention. Yet, understanding of the action space of AI agents and its constraints remains limited.

Existing studies capture usage patterns and development practices, but not aspects about the evolving ecosystem of tools for AI agents. So far, comprehensive overviews of the AI agent tool ecosystem are missing. These gaps motivate our approach to study the publicly available agent tool ecosystem via MCP servers. Whilst tool monitoring cannot replace the depth of platform studies or contextual richness of practitioner interviews, it offers complementary visibility into the action space of AI agents, and where agent usage may be concentrating. 

\begin{table*}[htbp]
\centering
\footnotesize
\setlength{\tabcolsep}{3pt}
\caption{The action space of AI systems measured across studies, by direct impact, generality, task domain, and geography}
\label{tab:action_space_comparison}
\resizebox{\textwidth}{!}{
\begin{tabular}{lrrrrrrrr}
\toprule
\textbf{Action space of} & \textbf{Workforce} & \textbf{Bing Copilot} & \textbf{ChatGPT} & \textbf{Claude.ai} & \textbf{OpenRouter API} & \textbf{Claude API} & \textbf{Comet} & \textbf{Public agent tools (here)} \\
\textbf{AI agents} & \textit{Global} & \textit{Microsoft} & \textit{OpenAI} & \textit{Anthropic} & \textit{$\sim$All providers} & \textit{Anthropic} & \textit{Perplexity} & \textit{$\sim$All providers} \\
 & (2023--2024) & (2024/01-09) & (2024/05-25/06) & (2025/11) & (2025/05-11) & (2025/11) & (2025/07-10) & (2024/11-26/02) \\
 & \textit{Workers \%} & \textit{Usage \%} & \textit{Usage \%} & \textit{Usage \%} & \textit{Token \%} & \textit{Usage \%} & \textit{Usage \% (User \%)} & \textit{Downloads \% (Tools \%)} \\
\midrule
\multicolumn{9}{l}{\textbf{Direct impact}} \\
Perception & --- & $\checkmark$ & $\checkmark$ & $\checkmark$ & $\checkmark$ & $\checkmark$ & $\checkmark$ & 31.7 (50.6) \\
Reasoning & --- & --- & $\checkmark$ & $\checkmark$ & $\checkmark$ & $\checkmark$ & $\checkmark$ & 3.3 (13.4) \\
Action & --- & --- & --- & --- & $\checkmark$ & $\checkmark$ & $\checkmark$ & 64.8 (36.0) \\
\midrule
\multicolumn{9}{l}{\textbf{Generality}} \\
Unconstrained environments & $\checkmark$ & $\checkmark$ & $\checkmark$ & $\checkmark$ & $\checkmark$ & $\checkmark$ & $\checkmark$ & 45.4 (14.3) \\
Constrained environments & $\checkmark$ & --- & $\checkmark$ & $\checkmark$ & $\checkmark$ & $\checkmark$ & $\checkmark$ & 54.6 (85.7) \\
\midrule
\multicolumn{9}{l}{\textbf{Task domain (SOC)}} \\
Computer \& Mathematical & 1.8 & 6.5 & 36.6 & 36.0 & 58.0 & 51.7 & 30.4 (28.8) & 92.0 (65.8) \\
Educational Instruction & 3.0 & 5.6 & 8.9 & 16.2 & 3.0 & 4.1 & 4.4 (5.0) & 0.3 (1.6) \\
Arts, Design \& Entertainment & 0.5 & 7.4 & 9.9 & 10.6 & 21.5 & 6.3 & 4.8 (5.3) & 0.3 (3.8) \\
Office \& Administrative & 4.8 & 13.1 & 7.9 & 8.3 & 3.0 & 15.1 & 5.0 (5.4) & 1.1 (2.9) \\
Life, Physical \& Social Science & 0.5 & 7.7 & 5.0 & 5.8 & 5.0 & 4.6 & 4.2 (4.5) & 1.1 (3.5) \\
Other & 89.4 & 59.6 & 31.7 & 23.0 & 9.5 & 18.1 & 35.6 (38.6) & 5.1 (22.5) \\
\midrule
\multicolumn{9}{l}{\textbf{Geography}} \\
US & 4.8 & $\checkmark$ & $\checkmark$ & 26.0 & 47.2 & $\checkmark$ & $\checkmark$ & 49.7 \\
China & 20.9 & --- & --- & --- & 6.0 & --- & --- & 11.0 \\
Europe & 6.7 & --- & $\checkmark$ & 20.5 & 21.3 & $\checkmark$ & --- & 18.4 \\
Rest of World & 67.7 & --- & $\checkmark$ & 53.4 & 25.5 & $\checkmark$ & --- & 20.8 \\
\bottomrule
\end{tabular}
}
\vspace{1mm}
\begin{minipage}{\linewidth}
\scriptsize
\textbf{Notes.} Rows are along the action space as defined in Section \ref{sec:characteristics}. The date ranges indicate the period of usage covered. SOC = U.S. Standard Occupational Classification (2-digit groups). Task domain rows report the share of activity associated with each SOC group. Tick marks indicate that a study includes usage data from an AI system in the specified action space, but a quantitative fraction is not available.

\medskip
Columns (left to right, sorted by degree of evidence on the action space of agents).

\textit{Workforce.} Workers~\% = global employment shares from ILO modelled estimates (ILOSTAT, Nov 2024; ISCO-08 1-digit, world aggregate), distributed to SOC 2-digit groups using BLS OEWS May 2023 proportions.

\textit{Bing Copilot.} Usage~\% = share of platform conversations mapped to SOC groups using Microsoft's IWA-to-SOC activity mapping \citep{yang2025adoption}. Because generic activities map to multiple occupations, the distribution is comparatively flatter.

\textit{ChatGPT.} Usage~\% = occupation shares approximated from Figure~12 of \citep{chatterji2025people}, global sample. ChatGPT is not available in China.

\textit{Claude.ai.} Usage~\% = Distribution of conversations by SOC, measured via CLIO system of a global sample \citep{anthropic2026aeiv4}. Claude.ai is not available in China.

\textit{OpenRouter API.} Token~\% = token share by task category \citep{aubakirova2025state}. Tasks classified via Google NL API and mapped to nearest SOC group. Roleplay (17\%) mapped to SOC 27 (Arts, Design \& Entertainment); General Q\&A (3\%) included in Other. OpenRouter geography from billing-region token shares.

\textit{Claude API.} Usage~\% = API usage shares by occupation category (global scope; no regional breakdown reported) \citep{anthropic2026aeiv4}.

\textit{Comet.} Usage~\% (User~\%) = share of queries (and adopting users) of the Perplexity's Comet agent. Original data is on O*NET career clusters, mapped to SOC groups using the Career Clusters crosswalk. \citep{yang2025adoption}

\textit{Public agent tools (here).} Downloads~\% (Tools~\%) = download-weighted tool usage share (npm/pypi) and share of all 177k tools from this study. Geography reflects download-weighted pypi shares across all tool types, geography of tools themselves not available.
\end{minipage}
\end{table*}

\section{Data}
\label{sec:data}

\subsection{Identification of MCP servers via Online Repositories}
\label{sec:online-repos}
\begin{figure}[htbp]
\centering
\includegraphics[width=\textwidth]{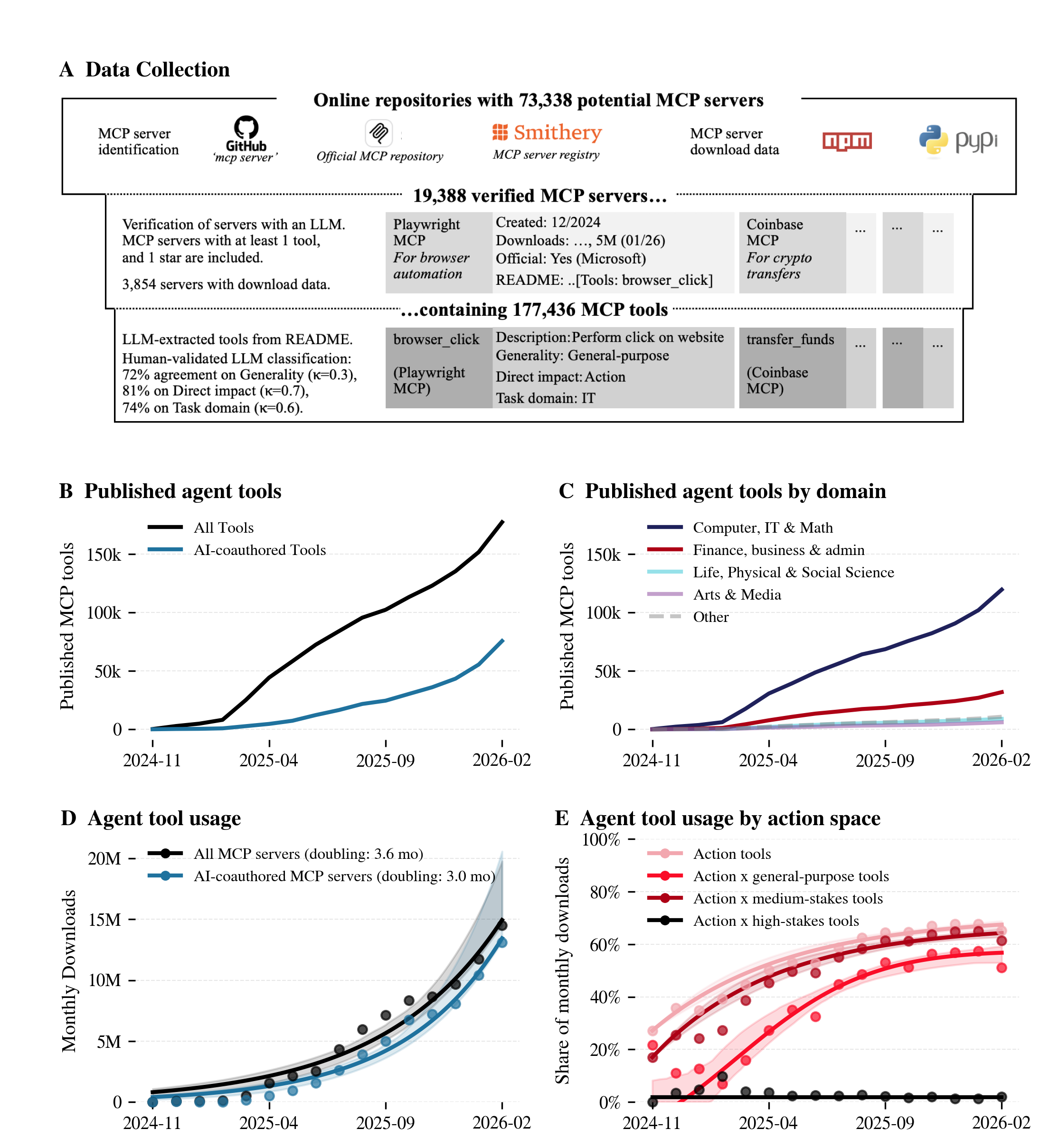}
\caption{\textbf{Monitoring 177k MCP tools}. \textbf{Panel~A} illustrates how we curate and classify 177k MCP tools from GitHub and Smithery using a human-validated LLM judge, along O*NET \citep{onetonlineONETOnLine} and US CAISI \citep{CAISI2025} taxonomies into generality (general-/narrow-purpose), direct impact (action/perception/reasoning), and task domain (Sections~\ref{sec:data} and~\ref{sec:methodology}). $\kappa$ is Fleiss' kappa across 14 expert validators ($n=100$ each; Appendix~\ref{app:validation}).
\textbf{Panel~B} shows cumulative MCP tools over time: all tools (black) by creation date and AI-coauthored tools (blue), by the month of first labelled AI evidence in an MCP repository (Method \& results in Sections ~\ref{sec:ai-detection-method} \& ~\ref{sec:ai-created-results}).
\textbf{Panel~C} shows cumulative tools by O*NET task domain. `Other' (dashed) aggregates remaining domains (Method \& results in Sections ~\ref{sec:topic-modelling} \& ~\ref{sec:res-task-domains}).
\textbf{Panel~D} shows monthly aggregate downloads of MCP server tools, proxying usage. Dots are monthly totals for all servers (black) and AI-coauthored servers (blue); downloads are attributed to the AI-coauthored series only from the date of first detected AI evidence. Lines show exponential fits $y = A\,e^{kt}$ (Nov~2024--Feb~2026). Legend reports the doubling time $\ln 2 / k$. Shading shows 95\% bootstrap confidence intervals.
\textbf{Panel~E} shows the share of monthly downloads by the type of tools, specifically the set of actions a tool allows an agent to take ('action space'). Dots are monthly shares. ``Action tools'' (light pink) denotes all tools classified as \emph{action}. Three subcategories cross-classify action tools by generality (``general-purpose,'' red) and O*NET occupational-impact stakes (``medium-stakes'' 50--75, dark red; ``high-stakes'' 75--100, black, on the 0--100 O*NET impact-of-decisions scale). Subcategory shares do not sum to the action total as dimensions are independent. Lines show WLS fits: asymptotic convergence for action and medium-stakes (95\% error-propagation CI), and poly-convergence for general-purpose and high-stakes (95\% bootstrap CI). Method \& results in Sections ~\ref{sec:direct-impact}, ~\ref{sec:generality}, ~\ref{sec:res-directimpact} \&  ~\ref{sec:res-generality}.}
\label{fig:methodology}
\end{figure}

Developers have been publishing MCP servers since November 2024 (when the protocol was released). Most are published on GitHub or on bespoke MCP registries. We identify MCP servers through 3 main data sources:

\begin{enumerate}
\def\labelenumi{\arabic{enumi}.}
\item
 \textbf{GitHub.} We searched for repositories with at least 1 star whose name, description, readme or tags include the string `mcp server' (n = 16,956 MCP servers in the final dataset).
\item
 \textbf{Smithery MCP registry} (n = 2,437 in the final dataset). We chose Smithery due to its permissive registry API and size, compared to other registries. For example, the official MCP registry is an order of magnitude smaller compared to Smithery, as of 02/2026 \citep{modelcontextprotocolOfficialRegistry}.
\item
 \textbf{MCP server lists on GitHub.} To ensure coverage of prominent MCP servers and identify `official' servers, we include servers on two popular MCP server lists: The official MCP repository \citep{githubGitHubModelcontextprotocolservers} (n = 841 in the final dataset), the most starred list `awesome MCP servers' \citep{archiveGitHubPunkpeyeawesomemcpservers} (n = 781 in the final dataset). All 1,366 of these also appear in source 1 or source 2.
\end{enumerate}

After deduplication and filtering out servers with 0 stars, we used an LLM (Claude Sonnet 4.5) to verify each entry was a valid MCP server rather than documentation or unrelated code. In addition, we only include servers with clearly defined tools in README files or descriptions. This process reduced the initial dataset from 73,338 potential servers to 19,388 verified servers. On these verified servers, we identified a total of 177,436 distinct agent tools as our final dataset.

We track the evolution of the MCP ecosystem by noting each server's creation date. Rather than analyzing only the latest February 2026 snapshot, we capture historical server versions: for servers created before October 2025, we collect READMEs and tools on 1 October 2025; for newer servers, we collect them on 1 February 2026. This approach, however, means that we do not track the changes of individual MCP servers.

\subsection{Identification of Official servers}
\label{sec:official-servers}
To verify that our analysis translates to AI agents integrated in production systems, we identified a subset of \textquotesingle Official\textquotesingle{} servers published by legally registered commercial entities, as marked on the MCP server lists, like PayPal, Stripe, Google or GitHub, providing 8,469 of 177k total tools. However, official servers are relatively more popular -- they comprise 45M of 78M total MCP server downloads on PyPI and NPM.

Official servers are built by the owners of the environments, like providers of APIs, for servers with narrow-purpose tools to access constrained environments. There are also official servers with general-purpose tools to access unconstrained environments, like Microsoft's official playwright server for web browsing, where the environments are not owned by the server creators. The official servers subset is representative of approximately 20\% of the UK AI sector, measured by AI-specific revenue produced by the entities creating these servers (The entities produce \textgreater3B GBP UK AI-specific revenue as of 2024, 10\% of entities with revenue data, \citep{undefinedArtificialIntelligence}). We confirm our findings with the subset of official servers (available upon request).

\subsection{Usage of MCP servers}
\label{sec:usage-mcp}

To estimate how popular MCP servers are, we tracked monthly download statistics (2024-11 to 2026-02) for the subset of MCP servers hosted on the Node Package Manager (NPM) and Python Package Index (PyPI) registries, which are the default for making local MCPs available. Figure \ref{fig:methodology} shows the data. We collect all usage data on March 1, 2026. We match 3,854 of 19,388 MCP servers (with 42,498 of 177,436 MCP tools) to download data. Package downloads serve as a proxy for ecosystem interest rather than a direct measure of runtime execution. This metric counts installation events (e.g., initialising an agent environment) rather than individual tool calls. Furthermore, it excludes private mirrors, cached installations, and usage via direct source code \citep{npmjsBlogArchive}. Consequently, our analysis focuses on relative usage trends and distribution shifts rather than absolute execution counts. A typical included use is the addition of an MCP server to a coding agent on a new virtual machine, e.g.: \texttt{claude mcp add playwright npx \textquotesingle\@playwright/mcp\@latest\textquotesingle{}} (NPM) or \texttt{uv tool install arxiv-mcp-server} (PyPI)

A typical non-included use is the addition of pre-mirrored MCP servers to chatbots, like adding a pre-verified connector to claude.ai. In addition, there might be remotely hosted servers or routine local workflows which do not require downloads, which are not included here. Thus, usage distributions should not be overinterpreted; the data might mostly indicate which tools are piloted most by developers rather than tools deployed in routine production workflows.

To assess ecological validity of download counts, we use Smithery use count data available as aggregates for uses of MCP servers made available via the Smithery CLI or OAuth (including remotely hosted servers), across 01-08/2025, see Appendix~\ref{app:topdown-domains}. We find that our sample is slightly biased towards developers (e.g. IT tools make up 90\% of PyPI/NPM MCP downloads vs. 80\% of MCP uses on Smithery).

\section{Methodology}
\label{sec:methodology}

\subsection{Assessing width of use across tasks}
\label{sec:topic-modelling}

To understand which domains are dominant in agent tools, we use a bottom-up and a top-down approach. 

\textbf{The bottom-up} approach uses a standard topic modelling pipeline BERTopic \citep{grootendorst2022bertopic}, to verify the top-down analysis, and to find naturally evolving sub-clusters within the main top-down categories. 
We use a pre-trained sentence transformer (Stella-400M) to embed each MCP server's title, description, and readme summary into 1024-dimensional vectors. We then apply UMAP dimensionality reduction, resulting in 5-dimensional vectors. Finally, we use HDBSCAN (min\_cluster\_size=0.3\% of dataset size, min\_samples=30\% of min\_cluster\_size, cluster\_selection\_epsilon=0.02) for clustering topics in dense embedding regions, while treating MCP servers in sparse embedding regions as outliers. We optimized the HDBSCAN parameters to maximize topic coherence while minimizing outliers, for a set range of 40-60 topics.

We validate the bottom-up clustering through two metrics, measured both for the main set and a held-out test set. We achieve a reasonable outlier rate of 25\% of topics not assigned to clusters (26\% in the test set), and a high topic coherence \citep{roder2015exploring} within clusters of 50\% (42\% in the test set). We name each cluster by prompting Claude Sonnet 4.5 to find a suitable name in the format "\textless2 words\textgreater{} tools", providing the ten most common terms in a cluster and five randomly sampled MCP server descriptions from the cluster (Visualisation in Appendix \ref{app:subclusters}).

\textbf{The top-down} approach ensures comparability to other AI usage studies \citep{gmyrek2023generative,brynjolfsson2018can,handa2025which,openaiChatGPTRelease}, by using the O*NET taxonomy for tasks and occupations of the US Department of Labor \citep{onetonlineONETOnLine}, to classify the main task covered by each tool. We use a three-level hierarchical classification approach originally proposed by Anthropic \citep[see Appendix][]{handa2025which}. We prompt the LLM to first allocate the tool into one of 12 high-level clusters, then one of the derivative mid-level clusters (n = 400 total), and then one of the associated O*NET tasks (n = 18796). This addresses the issue that all O*NET tasks combined do not fit in any LLM's context window (Appendix~\ref{app:hierarchy} for details, \ref{app:prompts} for the prompt). The matching of tasks to tools leads to 1+ tools for 2,060 tasks (10+ tools for 766 tasks). These 1874 tasks represent computer-based tasks (like filtered manually in \citep{shao2025future}). Manual validation by 14 humans holding Master's or PhD degrees in machine learning, each labelling n=100 tools, confirmed the accuracy of the LLM-based classification for the highest level (78\% agreement, Human-human Fleiss' $\kappa$ = 0.32, see Appendix~\ref{app:validation}). Assignment at the middle and bottom hierarchy levels is less reliable due to the extreme specificity of O*NET tasks and broader remit of many MCP tools. Thus, we focus on the highest hierarchical level for tasks (henceforth called `task domain').

For comparability to other studies, and consequentiality assignment we map tasks to occupations, using existing crosswalks from O*NET bottom-level tasks to the Standard Occupational Classification (SOC). Appendix~\ref{app:topdown-domains}, shows that this aggregation to SOC clusters is as reliable as our highest hierarchy level. In addition, we use the most common task and SOC clusters of all tools in a server to assign a server-level domain -- 84\% of servers have tools from only one task domain, and 74\% from only one occupation cluster. We use Claude.ai usage data for comparison \citep{handa2025which}. Claude.ai data is available for the same 12 highest level task domains for early 2025, and for occupational clusters also for November 2025 \citep{anthropic2026aeiv4}.

To assess risk, we mapped the amount of published tools to consequentiality of different tasks and occupations using O*NET impact data. The O*NET survey \citep{onetonlineWorkContext} identifies the consequentiality of certain occupations through a question that asks employees 'What results do your decisions usually have on other people or the image or reputation or financial resources of your employer?' By plotting tool availability against occupations that scored highly on this survey question, we can develop a proxy for whether agents are being used in 'high-stakes' settings.
\subsection{Assessing width of use across geographies}
\label{sec:geo-tools}

To understand where agents are used, we use country-level geographic splits of downloads of MCP servers, available for a subset of PyPI downloads, based on IP addresses downloading a package \citep{githubGitHubPypilinehaulcloudfunction}. 528 MCP servers with action tools have download data with geographical splits (2,467 of 11,174 MCP servers with action tools have download data at all). These 528 MCP servers accrued N=11.91M downloads for the covered period 2024-11 to 2025-10 (Data by geography is not yet available for 2025-11 to 2026-02). Geographic splits are not available to us for NPM.
\subsection{Assessing perception vs. reasoning vs. action}
\label{sec:direct-impact}

\begin{table*}[htbp]
\centering
\caption{Direct impact and functionality of Agent tools.}
\small
\resizebox{\textwidth}{!}{%
\begin{tabular}{lll}
\toprule
\textbf{Direct impact} & \textbf{Functionality} & \textbf{Examples} \\
\midrule
Perception & Sensors & Internal database, monitoring, diagnostics, GUI, voice, internet search, physical world \\
\midrule
Reasoning & Planning & Task-decomposition, path-finding models \\
 & Analysis & Scratchpads, calculators, simulations \\
 & Resource mgmt. & Memory, self-management \\
\midrule
Action & Authentication & Login, CAPTCHA, wallet \\
 & Computer use & Application-specific GUI interaction, website interactions, computer use \\
 & Running code & Sandboxed code interpreter, file operations, code execution \\
 & Software extensions & Calendar, social media API \\
 & Physical extensions & Robotic arm, laboratory tools in factory setting, robot in an open environment \\
 & Human interaction & Phone calls \\
 & Agent interaction & Multi-agent workflows, third-party agent interactions \\
\bottomrule
\end{tabular}%
}
\textit{Notes:} Taxonomy from \citet{CAISI2025}, GUI = Graphic user interface.
\label{tab:direct-impact}
\end{table*}
To understand whether agents observe and access or actively modify the digital economy, we classify direct impact and functionality of agent tools (see Table~\ref{tab:direct-impact}). Following a recent CAISI \citep{CAISI2025} taxonomy, we use Claude Sonnet 4.5 to classify direct impact as action vs. reasoning vs. perception tools (Prompt in Appendix~\ref{app:prompts}). Human validators (n=14), each labelling 100 tools agree 81\% (Fleiss' $\kappa$=0.7) with Sonnet 4.5's direct impact classification, 85\% (Fleiss' $\kappa$=0.5) on functionality conditional on matching direct impact (Appendix~\ref{app:validation}). To identify time trends, we use weighted least squares (WLS) regressions throughout the work, weighting by downloads (if not indicated otherwise in the figure). Regressions are indicated in the figures, and the linear, quadratic or asymptotic regression specifications are chosen to balance simplicity and explanatory power. We assign the direct impact classification on tool level, as one MCP server may have tools to access and tools to modify external environments. However, a minority of servers does not have action tools. 

\subsection{Assessing narrow-purpose (constrained) vs. general-purpose (unconstrained) access to external environments}
\label{sec:generality}

\begin{table*}[htbp]
\centering
\caption{Generality of AI agent tools (adapted from \citet{CAISI2025})}
\small
\begin{tabular}{ll}
\toprule
\textbf{Generality} & \textbf{Examples} \\
\midrule
Narrow-purpose (constrained environment) & Access to software or platforms via API, data retrieval \\
General-purpose (unconstrained environment) & Deep research, browser use, computer use \\
\bottomrule
\end{tabular}
\label{tab:generality}
\end{table*}
To understand whether agents access and modify constrained or unconstrained environments like the web, we use Claude Sonnet 4.5 to classify agent tool generality. We distinguish between \emph{narrow-purpose} tools, and \emph{general-purpose} tools, such as web browsing (see Table~\ref{tab:generality}). 
Human experts (n=6), each labelling 100 servers, show agreement of 72\% (\emph{Fleiss'} \(\kappa\)\emph{=0.3}, fair agreement) with Claude Sonnet 4.5 on the generality label. 

One MCP server typically bundles tools to interact with one environment. Generality is a property of the environment an MCP server provides access to, and all tools on a MCP server typically interact with the same environment. Thus, we classify generality on server-level, and assign the same generality to all tools of a server.

\subsection{Assessing AI assistance in the creation of MCP servers}
\label{sec:ai-detection-method}

To understand the potential for AI agents expanding their own action space, we identify whether an MCP server has been created with the help of agent systems. The ability of AI agents to create their own tools may have two opposing effects on our data of public agent tools. There might be more AI agent tools, since tool creation is less blocked by human effort. There might be less \textit{public} AI agent tools, if AI agents can create reliable tools on-the-fly tailored to the needs of a particular task.

We identify AI-created MCP servers through four categories of evidence in each repository's GitHub metadata: (i) \texttt{Co-Authored-By} commit trailers referencing known AI coding agents (e.g.\ Claude Code, GitHub Copilot), (ii) AI tool configuration files (e.g.\ \texttt{CLAUDE.md}, \texttt{.cursorrules}), (iii) commits or pull requests from known AI bot accounts (e.g.\ \texttt{copilot[bot]}), and (iv) explicit mentions of AI tool names in commit messages or pull request bodies (e.g.\ \texttt{@codex}). A server is classified as AI-created if any single piece of evidence from any criterion is found. For each repository, we scan the full commit history (up to 10,000 commits), the 30 most recent pull requests, and the complete file tree via the GitHub REST API. This identification approach only captures clearly labelled footprints of AI agents, underestimating actual AI assistance (see validation details in Appendix~\ref{app:ai-detection}). We also compute a first-month restricted variant that only considers evidence from within 30 days of repository creation, reducing false positives from AI tools adopted after initial development (see Appendix~\ref{app:ai-detection}) to analyse time trends accurately. For timeline figures (Figure~\ref{fig:methodology}, panels~b and~d), we count each server's tools and downloads as AI-coauthored only from the month of the first detected AI evidence in its commit history, rather than from the server creation date. This ensures that servers which adopted AI tooling after initial development are not retroactively counted as AI-coauthored for earlier periods.

\section{Results}
\label{sec:results}
\subsection{How widely are AI agents used across task domains?}
\label{sec:res-task-domains}

We find tools designed for software development and IT tasks account for 67\% of the total dataset. 90\% of downloaded MCP servers mainly hosted software development and IT tools (Table \ref{tab:domains}, third column). This concentration suggests that the primary current utility of agents is to accelerate technical workflows rather than to automate broader economic tasks. 18\% of tools support finance and business management tasks (5\% of MCP server downloads).

\begin{table*}[htbp]
\centering
\caption{Agent tools by task domains}
\small
\setlength{\tabcolsep}{3pt}
\rowcolors{2}{rowgray}{white}
\begin{tabular}{>{\raggedright\arraybackslash}p{3.5cm}>{\raggedright\arraybackslash}p{2.2cm}>{\raggedright\arraybackslash}p{1.7cm}>{\raggedright\arraybackslash}p{1.3cm}>{\raggedright\arraybackslash}p{2cm}>{\raggedright\arraybackslash}p{2.3cm}}
\toprule
\rowcolor{white}
\textbf{Task Domains} & \textbf{MCP Servers published}\newline\textbf{(\%, downloads \%)} & \textbf{Tools published} \textbf{(\%, downloads \%)} & \textbf{Claude.ai}\newline\textbf{(usage \%)} & \textbf{Bottom-up}\newline\textbf{subclusters} & \textbf{Examples} \\
\midrule
Design, implement, and maintain diverse IT systems & 12,004\newline(68\%, 90\%) & 119,685\newline(67\%, 94\%) & 52\% & Search, Security, \ldots & context7, github-mcp-server \\
Business management, finance, and customer service & 2,397\newline(14\%, 5\%) & 31,882\newline(18\%, 4\%) & 11\% & Trading, E-Commerce, CRM, SEO & excel-mcp-server, mcp-boilerplate \\
Conduct scientific research and technical analysis & 1,273\newline(7\%, 3\%) & 8,989\newline(5\%, 1\%) & 6\% & ChemBio, Math, Weather & ghidramcp, deep-research \\
Create and preserve art, culture, and religious artifacts & 723\newline(4\%, $<$1\%) & 6,053\newline(3\%, $<$1\%) & 15\% & Image, Music & minimax-mcp, elevenlabs-mcp \\
Manage education, HR, and professional development programs & 201\newline(1\%, $<$1\%) & 1,857\newline(1\%, $<$1\%) & 8\% & - & anki-mcp-server, mcp-server \\
Other* & 931\newline(5\%, 1\%) & 8,968\newline(5\%, $<$1\%) & 6\% & - &  \\
\bottomrule
\end{tabular}
\rowcolors{1}{}{}
\vspace{1mm}
\begin{minipage}{\linewidth}
\scriptsize
*Regulatory enforcement, public safety, industrial, logistics, sustainability, healthcare tasks.
Sorted by server downloads \%. Download data is on server level, allocated assuming 1 server install = 1 use of every tool on the server. Claude.ai usage \% from Anthropic~\citep{handa2025which}. Bottom-up subclusters from Appendix~\ref{app:subclusters}. See Section~\ref{sec:topic-modelling} for top-down classification methodology.
\end{minipage}
\label{tab:domains}
\end{table*}

\begin{figure}[htbp]
\centering
\includegraphics[width=\textwidth]{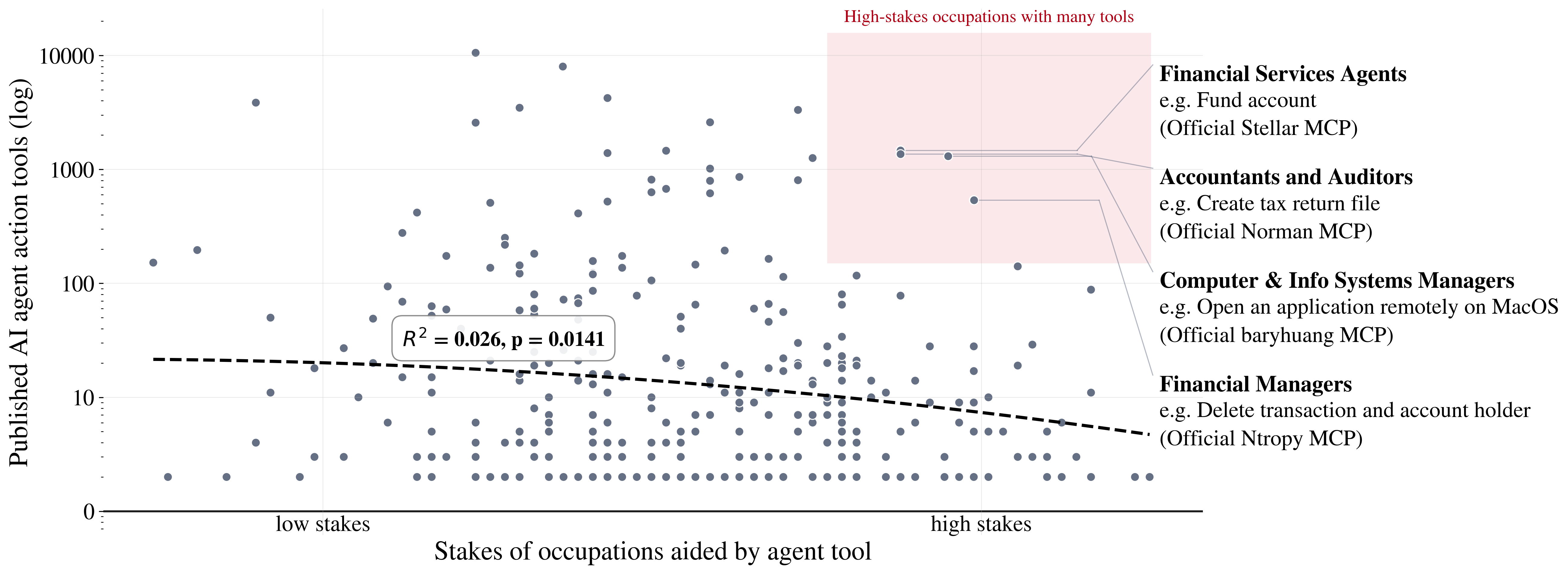}
\caption{\textbf{Consequentiality distribution of AI agent actions.} The figure shows the stakes of computer-based occupations, and the number of tools related to each occupation. Each dot represents one SOC-O*Net occupation. Occupation stakes are based on an O*NET survey \citep{onetonlineONETOnLine} asking employees to rate `What results do your decisions usually have on other people or the image or reputation or financial resources of your employer?' on a scale of 0-100. Absolute values are meaningless and imprecise, thus the axis labels are omitted. The y-axis (log scale) shows the number of published AI agent action tools mapped to each occupation. The dashed curve shows a quadratic polynomial fit. The fit explains little of the cross-occupation variance ($R^2 \approx 0.03$; an F-test rejects that all slope coefficients are jointly zero, $p = 0.015$), reflecting substantial heterogeneity. The pink-shaded region highlights high-stakes occupations (score $>$75) with many tools. Occupations without any associated agent tools -- near-exclusively non-computer-based occupations -- are excluded.}
\label{fig:stakes-occupations}
\end{figure}

Are AI agents used in consequential settings? We classified action tools as supporting high- or low-stakes tasks (as judged by the O*NET classification system). We find most action tools support medium-stakes occupations such as computer systems administration, with relatively few tools for low-stakes or high-stakes tasks (Figure ~\ref{fig:stakes-occupations}). However, finance represents a significant outlier: high-stakes financial occupations have  disproportionately more action tools than predicted by the overall pattern. 
Beyond finance, we identified relatively few official servers enabling high-stakes integrations. Existing high-stakes MCP servers include medication management, tax filing, drone navigation, legal document generation. Within high-stakes occupations, tools typically support 
lower-stakes subtasks - for example, medical tools enable image processing but not prescription authorisation. This represents a lower bound, as higher-stakes tools may exist in private deployments.

\subsection{How widely are AI agents used across geographies?}
\label{sec:res-geo}
\begin{figure}[htbp]
\centering
\includegraphics[width=\textwidth]{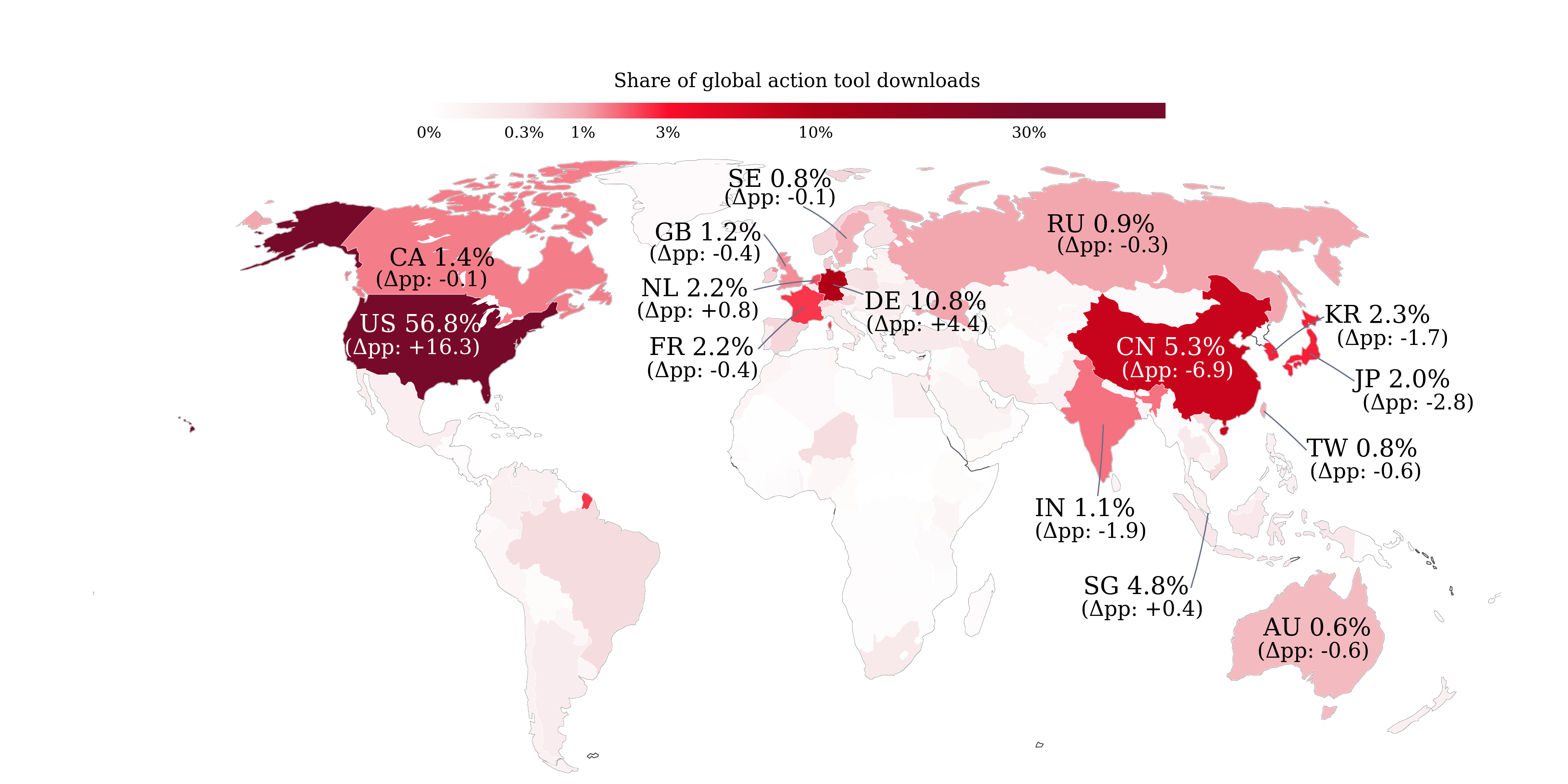}
\caption{\textbf{Geographic distribution of AI agent actions.} Share of worldwide PyPI downloads of MCP servers with action tools 11/2024 to 10/2025 (colour intensity), in brackets percentage point change of share H1 to H2 2025. $N{=}6.73$M downloads of 528 MCP servers with action tools have download data by geography (of 11,174 MCP servers with action tools, 2,467 have download data; see Section~\ref{sec:geo-tools}). CA (Canada), US (United States), SE (Sweden), GB (Great Britain/United Kingdom), NL (Netherlands), DE (Germany), FR (France), KR (South Korea), CN (China), JP (Japan), TW (Taiwan), IN (India), SG (Singapore), AU (Australia).}
\label{fig:geographic-distribution}
\end{figure}
Analysis of IP address data from package registry downloads indicates that agent deployment is heavily concentrated in the United States, which accounts for half of global downloads in 2025 (Figure~\ref{fig:geographic-distribution}). Western Europe follows with approximately 20\%, while China accounts for 5\%, Singapore for 5\% and Korea for 2.3\%. Other countries or regions account for less than 2\% each. We note that this distribution likely reflects the Western-centric user base of the PyPI registry and may underrepresent activity in regions utilizing alternative distribution channels.

\subsection{What fraction of AI agents are used for perception, reasoning or action?}
\label{sec:res-directimpact}

Tools are increasingly enabling agents to modify the environment, with more tools and tool downloads for action tools over time. Our longitudinal analysis reveals a marked transition in agent capabilities from passive observation to active environmental modification.

\begin{figure}[htbp]
\centering
\includegraphics[width=\textwidth]{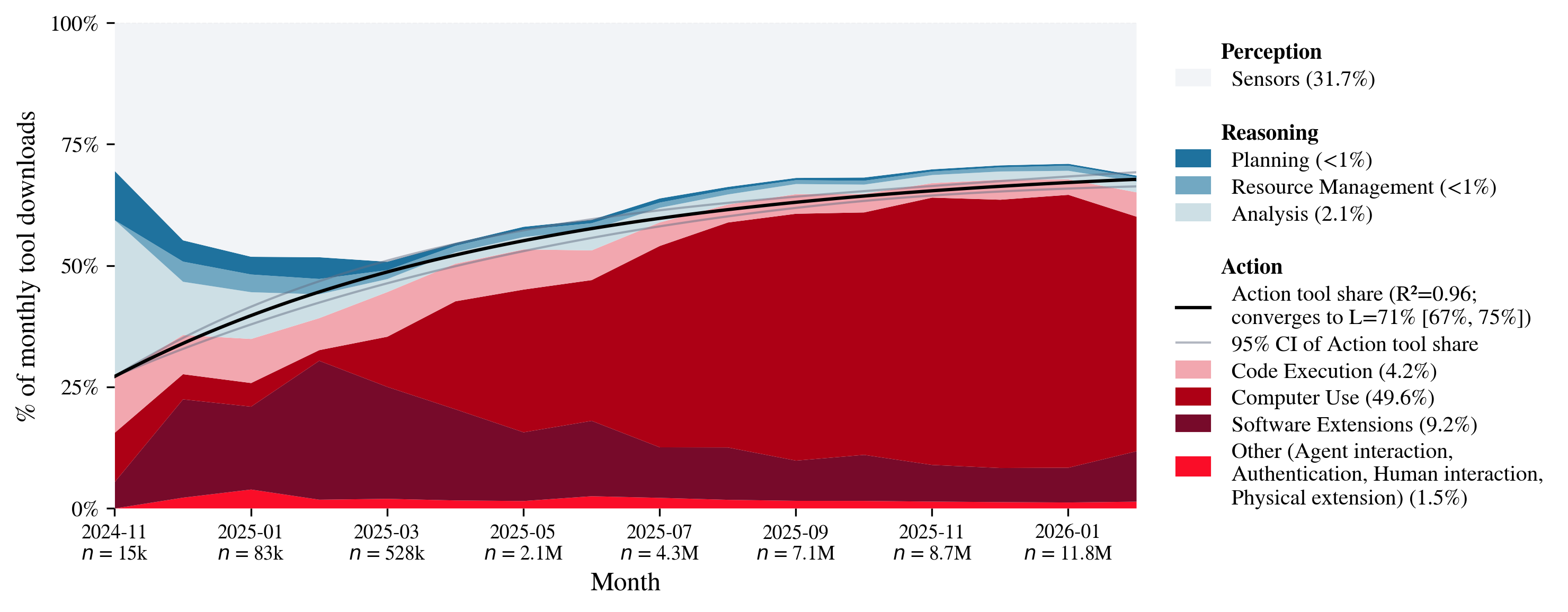}
\caption{\textbf{AI agent tool usage for perception, reasoning and action.} Stacked area chart showing the percentage of monthly tool downloads on PyPI and NPM (y-axis) by tool functionality subcategory, from Nov~2024 to Jan~2026 (x-axis; $n$ indicates total monthly downloads). Stacked areas are grouped into three direct impact types (bottom to top): \emph{Action} (red shades), \emph{Reasoning} (blue shades), and \emph{Perception} (grey). Parenthetical percentages in the legend show each subcategory's overall download share across all months. Subcategory definitions follow the taxonomy in Section~\ref{sec:direct-impact}; software extensions are tools for specific software packages and APIs, code execution covers command-line tools (e.g., a bash tool), and computer use includes tools for mouse-based computer control, browser automation, and GUI interaction. The black trend line shows an asymptotic convergence model $y(t) = L - (L - y_0)\,e^{-kt}$ fitted via weighted least squares (by monthly downloads). The asymptotic limit $L$ and 95\% confidence interval (from the parameter covariance matrix) are on the legend; grey lines show 95\% CI of overall trend. LLM classification validated by human experts (78\% agreement, see Appendix~\ref{app:validation}). Download data is on server level, allocated assuming 1 server install $=$ 1 use of every tool on the server.}
\label{fig:action-tools-trend}
\end{figure}

\textbf{Action tool usage growth:} Action tools rose from 27\% to 65\% of downloads over the 16-month period (Figure~\ref{fig:action-tools-trend}). This shift from perception tools (which allow agents to observe environments) to action tools (which enable direct environmental modification) was driven primarily by adoption of general-purpose tools for browser use and computer control.

\textbf{Commercial adoption:} The trend from perception to action is particularly acute among tools released by registered commercial entities, where the download share of action tools increased from 21\% to 71\% over the same period. This shift is driven largely by the adoption of general-purpose \textquotesingle computer use\textquotesingle-type tools, including tools like playwright for browser automation, remote mobile phone use and AppleScript-desktop integration.

With the introduction of MCP servers in 2024, initial tool downloads focused on reasoning and perception tools. The rise of action tools was initially driven by specific software extension tools and to a small extent by code execution tools. Increasingly, general computer use tools gain most downloads.

\subsection{What fraction of AI agents are used to access narrow, constrained environments or general, unconstrained environments?}
\label{sec:res-generality}

\begin{figure}[htbp]
\centering
\includegraphics[width=\textwidth]{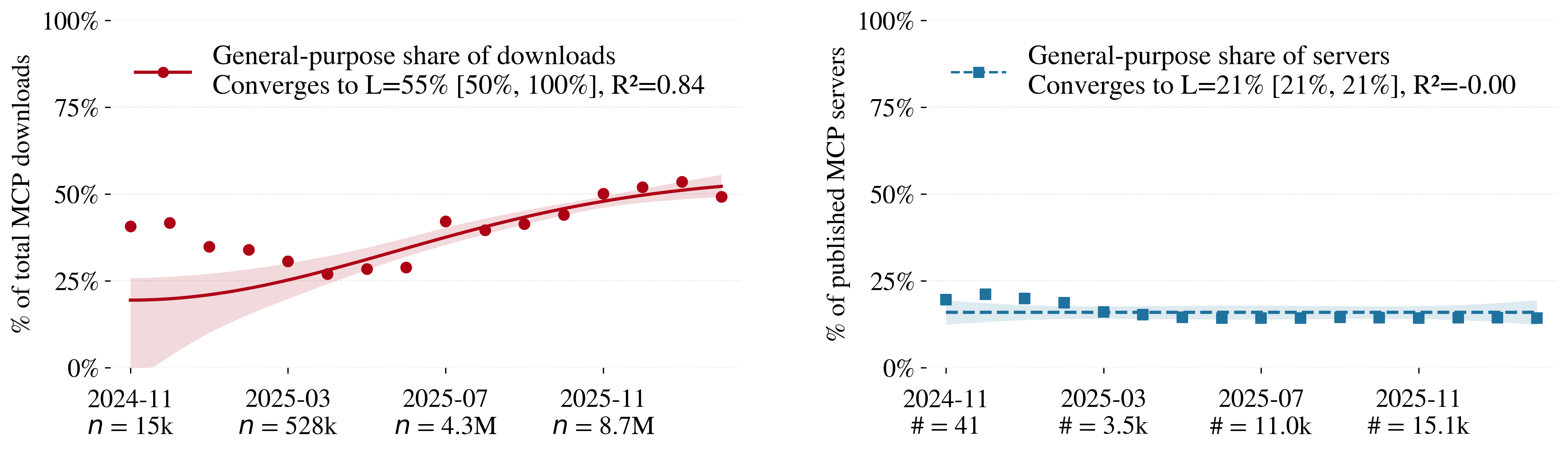}
\caption{\textbf{General-purpose tool share over time.} \textbf{Top}: general-purpose share of total monthly downloads ($n$~=~total npm/PyPI downloads per month); \textbf{bottom}: general-purpose share of cumulative published servers (\#~=~cumulative server count). Dots are observed monthly values. Curves: polynomial-convergence model $y = L - \exp(a + bt + ct^2)$ fitted via WLS. Shading: 95\% confidence intervals---wild bootstrap (top) and standard covariance-based (bottom). Download-weighted general-purpose share converges toward $L = 55\%$ [50\%, 100\%] ($R^2 = 0.84$); the count-based share remains stable near $L = 21\%$ [21\%, 21\%] ($R^2 \approx 0$).}
\label{fig:generality-tools}
\end{figure}

General-purpose tools (operating in unconstrained environments like the open web) grew from 41\% to 50\% of downloads
(Figure \ref{fig:generality-tools}). This shift was concentrated in action tools: 94\% of general-purpose server downloads involved action capabilities (e.g., browser automation,
arbitrary code execution), while perception tools remained predominantly narrow-
purpose (95\% of downloaded perception tools operate in constrained environments, e.g., accessing data via APIs).
This correlation suggests that potentially consequential agent actions are currently occurring in the least controlled environments (e.g., an agent browsing the web or using a computer) rather than in restricted, secure API integrations. 

\subsection{What fraction of AI agent tools are created with the support of AI agents?}
\label{sec:ai-created-results}

As shown in Figure~\ref{fig:methodology}(b) and~(d), AI-coauthored tools represent a substantial and growing segment of the MCP ecosystem. The cumulative count of AI-coauthored tools (blue line in panel~b) tracks the overall tool growth closely, and AI-coauthored server downloads (blue in panel~d) follow a similar trajectory to total downloads.

\begin{figure}[htbp]
\centering
\includegraphics[width=\columnwidth]{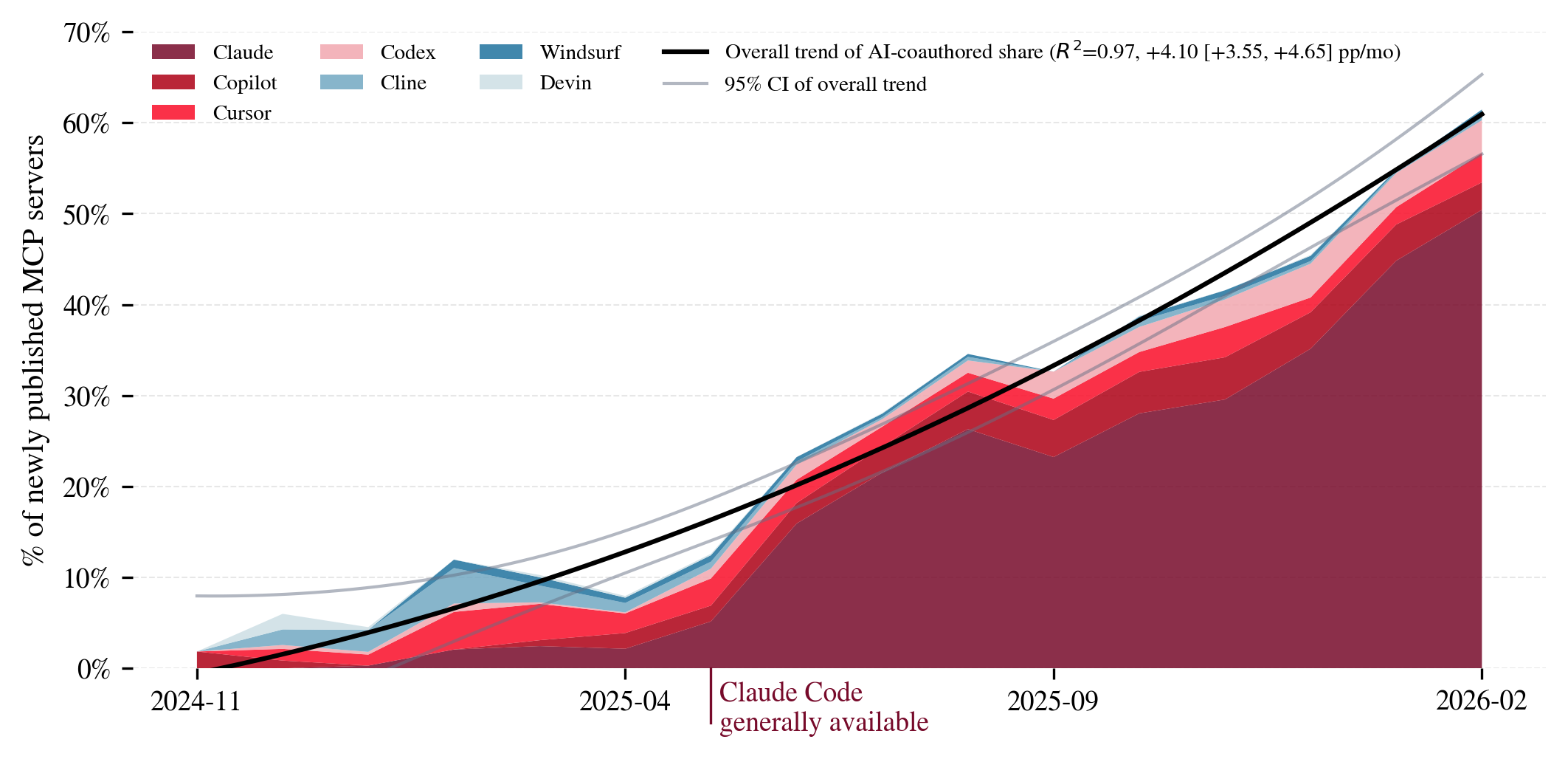}
\caption{\textbf{AI-coauthored MCP servers by AI coding agent.} Monthly share of newly published MCP servers with AI assistance detected in first-month commits, by coding agent (stacked area). The trend line shows a WLS quadratic best fit weighted by monthly server count ($R^2$=0.97, average marginal change +4.10 [+3.55, +4.65] pp/mo, 95\% CI via delta method); grey lines show 95\% CI. Claude dominates AI-assisted MCP server creation (69\% of AI-coauthored servers).}
\label{fig:ai-created-results}
\end{figure}

We detect AI assistance in first-month commits for 5,494 of 19,388 MCP servers (28.3\%) and 64,489 of 177,436 tools (36.3\%). The share of newly created MCP servers with detected first-month AI assistance rose from 6\% (01/2025) to 62\% (02/2026). Figure~\ref{fig:ai-created-results} shows the monthly share of AI-coauthored servers by coding agent. Claude dominates AI-assisted MCP server creation (3,770 servers, 68.6\% of AI-coauthored servers), followed by Cursor (507, 9.2\%), Copilot (502, 9.1\%), and Codex (328, 6.0\%; combining ChatGPT and Codex, both OpenAI products).

\subsection{High-stakes example: Anticipating autonomous transaction trends via MCP tools}
\label{sec:res-highstakes}

Financial authorities and regulators are aiming to `understand better how AI adoption and use cases -- such as agentic AI -- are transforming the wider economy' \citep{bankofenglandBankEnglands} and especially payment systems \citep{fcaSprintSummary}. We demonstrate the potential of monitoring agent tools to identify early AI agent payment trends that could inform financial authorities, jointly with other indicators such as scraping websites of technology companies or product announcements \citep{undefinedArtificialIntelligence}, sectoral surveys \citep{bankofenglandArtificialIntelligence} or agent system usage data.

\begin{figure}[htbp]
\centering
\includegraphics[width=\textwidth]{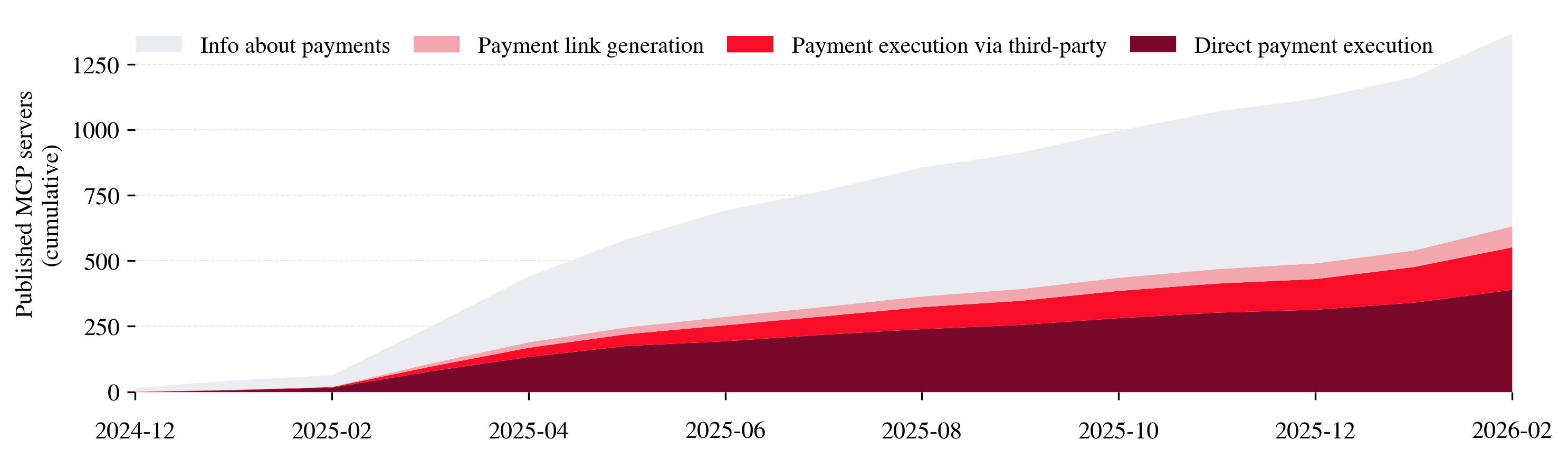}
\caption{\textbf{Agentic transaction tools by autonomy level.} The figure shows the number of publicly available MCP servers (\#) with any tools for payments in a given month. Human validators agree with the LLM classification 83\% (n=6, Fleiss' $\kappa$ = 0.42, Prompt uses exact wording of legend and examples, see Appendix~\ref{app:prompts}).}
\label{fig:agentic-payments}
\end{figure}

MCP tool creation trends show early signs of autonomous payments. One risk that financial regulators are concerned with is whether agents may enable higher risk transactions like cryptocurrencies with less regulatory oversight, less reversal options and at a greater scale. Using MCP tool monitoring, we find a trend towards tools enabling direct agent payment infrastructure for cryptocurrencies.~This may exacerbate risks to systemic financial stability \citep{von2024crypto}. In line with Appendix~\ref{app:monitoring}, deeper monitoring of agent tools for accessing the financial system could inform targeted requests for interviews and usage data investigation.

Figure~\ref{fig:agentic-payments} shows the growth of MCP servers with payment execution capabilities, rising from 47 servers in January 2025 to 1,578 servers in February 2026, with corresponding growth in downloads over the same period.

\section{Discussion: What do these tool use findings tell us about the risks of AI agents in different domains?}
\label{sec:discussion}
We find that the action space of AI agents is increasing rapidly: Agents in early 2026 have access to 36x more public tools than a year before (from approximately 4,888 tools in 01/2025 to 177,000 tools in 02/2026). Usage of these tools, measured by server downloads, increased by two orders of magnitude in the same period (0.08 to 14 million). Usage increasingly concentrates on general-purpose tools that enable agents to take action in unconstrained environments, like on the internet. While we do not measure risks directly, our findings may have implications for managing and governing AI agents' risks in the future, in particular: 

First, the rise in deployment of AI agents, in particular with general-purpose tools, expands the attack surface exposed to malicious actors. Misuse and security risks, like prompt injection for credential theft become more potent when agents can execute code and access file systems.

Second, expanding action spaces amplify the consequences of misalignment or errors. When agents modify external environments with little oversight, mistakes might propagate. As agents handle tasks with real-world impact, the consequences of releasing misaligned updates to models on which many agents depend increase. 

Third, the number and usage of tools for software development and financial tasks is particularly high. In these domains, the scale and speed of agentic action might cause structural changes. Indeed, some changes are already visible in the job market for entry-level software developers \citep{brynjolfsson2025canaries}. Tracking public agent tools provides measurable early signals of deployment patterns, to anticipate risks and prioritise further monitoring. However, this approach must be accompanied with data on private, internal agent tools, and approaches to measure usage context of general-purpose tools. This method will cease to be useful if AI agents are building their own tools as they need them. 

Fourth, the shift toward general-purpose tools complicates tool-based governance to reduce individual risks. Narrow-purpose tools are easier to govern than general-purpose tools: a cryptocurrency transfer tool has a clear risk profile and clear use cases. A browser use tool can download both necessary files and, in some cases, malware. If general-purpose tools grow in popularity, the user review necessary for dual-use tool calls becomes arduous. Current agent systems like Claude Code's settings.json permit or block specific (narrow) tools, while requiring user review for potentially risky general-purpose tools like fetching data from websites. Future agent systems could condition permissions on the context of general-purpose tool calls. For consequential actions -- large financial transfers, legal registration etc. - developers and regulators could require human authentication.

Fifth, the increasing dominance of action tools for general, unconstrained environments may increase structural risks across domains. Potentially consequential agent actions are  increasingly occurring in the least controlled environments (e.g., an agent browsing the web or using a computer) rather than via restricted, secure APIs. AI agents might permeate unconstrained environments designed for human use, while not being easily differentiable vs. human users. For such scenarios, scholars have proposed agent IDs \citep{chan2024ids}.

In Appendix~\ref{app:monitoring} we compare when government bodies might want to use monitoring of agent tools, like MCP monitoring, and when web scraping, interviews or usage data reviews. Tool monitoring is particularly helpful for answering early, explorative questions on use, in the absence of clear critical public use cases. Later, surveys, interviews and usage data can be used to deepen the understanding of agent use cases. 

\section{Conclusions and future work}
\label{sec:conclusions}

\subsection{Conclusion}

By analysing 177,436 agent tools and tracking their downloads over a 16-month period, this study provides the first systematic measurement of the action space of AI agents---the set of actions agents can take in external environments through their tools. We find that the action space of AI agents expanded significantly across the five dimensions that define it: Tool availability increased by more than an order of magnitude, from approximately 4,888 tools in 01/2025 to 177,000 tools in 02/2026. The share of \emph{action tools} (those that modify environments rather than merely perceive them) rose from 27\% to 65\% of downloads (21\% to 71\% for tools built by registered companies). \emph{general-purpose tools} enabling access to unconstrained environments grew from 41\% to 50\% of downloads, with 94\% of these involving action capabilities. Tools predominantly support medium-stakes occupations in software development (67\% of tools, 90\% of downloads), with little but some expansion into higher-stakes domains, like for financial transactions and cryptocurrency. Tools are deployed across \emph{geographies} but remain heavily concentrated, with the United States accounting for 57\% of downloads, followed by Western Europe (20\%) and China (5\%). A significant and growing share of tools are \emph{AI-coauthored}: 28\% of MCP servers (36\% of tools) show evidence of AI assistance, with the share of newly created AI-coauthored servers rising from 6\% (01/2025) to 62\% (02/2026).

These patterns indicate the action space of AI agents is expanding significantly and unevenly: agents are transitioning from primarily observing environments to actively modifying them, increasingly through general-purpose tools in unconstrained environments, with early deployment in high-stakes domains.

\subsection{Limitations and future work}

MCP repositories are just one way in which developers distribute tools for AI agents, and our analysis of these repositories may not capture the full breadth of the action space available to agents in practice. Developers may build custom integrations, use proprietary internal tooling, or distribute tools through channels not covered by our data sources. As such, our results represent a lower bound on the actions available to AI agents, rather than aiming to be comprehensive. Future agent monitoring work could extend this paper in several ways:

\textbf{Multi-level monitoring expansion.}~Our tool-level analysis provides insight into the tools that shape agent action spaces, but misses agent use context. Future work should expand to \emph{agent systems} and their usage to track the orchestration of agent tools including autonomy; and \emph{agent actions} measuring actual agent actions in external systems, like on GitHub, on the internet, on digital markets, on payment flows etc.

\textbf{Sector-specific follow-ups.}~ Finance and agent payment monitoring requires a framework to monitor and mitigate potential systemic agent risk for financial stability. This could track: (1) agent actions as share of total transactions; (2) concentration in specific market segments or time windows; (3) correlation patterns; (4) early indicators of payment system stress or market instability attributable to agents. Such frameworks are needed for other high-stakes domains as agent adoption grows.

\textbf{Risk scenarios and thresholds.}~Connecting measurement to societal-scale risks requires sector-specific risk models. For example, what concentration of agent transfers creates cascading failure vulnerabilities? These models should identify thresholds where agent activity transitions from contained to systemic risk, informing when additional mitigations become necessary.

\textbf{Methodological \& validation improvements.}~As a growing majority of tools fall under \emph{action}, future monitoring requires an updated and more granular taxonomy beyond three levels of direct impact (e.g. \citealt{kasirzadeh2025characterizing}). The current approach captures the 2024-2026 shift from perception to action tools; a future approach might differentiate between the duration, reversibility, consequentiality or modification degree of actions, which might need to be sector specific. More precise mapping of economic tasks is required to make more granular claims on agents' potential economic impact, e.g. with validations of O*NET experts. The broad classification of `official servers' could be dissected into specific action environments of agents, and their share of economic activity by subsector. Future work could also include scanning for actively hosted MCP servers on the web \citep{knosticFindServer}, and traces of other agent protocols such as the Agent Payments Protocol.

\textbf{General-purpose tools tracking via specific-purpose skills.} The approach used here can track the prevalence of general-purpose vs. narrow-purpose tools, but cannot meaningfully track and classify actions done with general-purpose tools. If the trend towards general-purpose tools continues, tracking tools will be increasingly uninformative on sector-specific rollouts. However, general-purpose tools will still require specific orchestration, which could be monitored via \emph{skills} or AGENTS.md \citep{agentsAGENTSmd} published in online registries. Skills are specific workflows that integrate tools, and thus can be more task- and sector-specific.

\section*{Acknowledgements}

For careful review, thank you to Alan Chan, Stephen Casper, Elliot Jones, Toby Pilditch, Shahar Avin, Catherine Fist, Kimberly Mai, Roxana Radu, Georgiana Gilgallon, Karina Kumar, Mahmoud Ghanem and Neil Perry. Thank you to Chris Summerfield and Andrew Strait who provided extensive advice throughout the project. Thanks for initial brainstorming go to J.J. Allaire, Kola Ayonrinde, Sid Black and the Societal Resilience team at UK AISI. This agent tool monitor has been part of an ongoing collaboration between the UK AI Security Institute and the Bank of England \citep{bankofenglandBankEnglands}. Thank you to the joint work with Elliot Jones and Andrew Walters (Bank of England), as well as Rosco Hunter and Ture Hinrichsen (AISI) to make the monitor useful for governmental foresight on agent deployments.

Beyond the mentioned use of LLMs for classification, the authors used Claude Code and Claude Opus 4.5 and 4.6 for coding, minor edits, and formatting.

\bibliographystyle{ACM-Reference-Format}
\bibliography{paper}

\appendix

\section{Supplementary Materials}

\subsection{Detailed human validation results}
\label{app:validation}

\textbf{Human validators (n=14) agree} \textbf{78\% with Claude Sonnet 4.5's O*NET task classification}. We validated the findings of the LLM classifications with graduates with ML Master's or PhD degrees, sourced from Prolific (n=6 annotators classifying a random subset of 100 MCP servers, and n=8 annotators classifying a random subset of 100 MCP tools). To rigorously assess internal validity, our human validation approach differs significantly from previous non-blind approaches where validators judged if the LLM `assigned {[}an item{]} to an acceptable task' \citep[see][supplementary materials]{handa2025which}. Instead, we use a blind approach, with validators assigning items (MCP tools or MCP servers) to occupational tasks at the highest hierarchy level, and compare assignments. This yields 78\% agreement (Fleiss' $\kappa$ = 0.32) on MCP server level with the LLM classification (server-level classification derived as the mode of LLM tool level classification), and 74\% agreement (Fleiss' $\kappa$ = 0.57) on MCP tool level. Assignment at lower hierarchy levels is less reliable due to the extreme specificity of O*NET tasks and broader remit of many MCP tools (e.g. the agreement between GPT-5 and Sonnet 4.5 at \textless70\%), thus we focus on aggregated statistics on the highest levels for tasks, and on occupations.

\textbf{Human validators (n=14) agree 81\% with Sonnet 4.5's direct impact classification}. The majority of MCP servers contain tools of different levels of direct impact, such as read\_file (perception) and edit\_file (action), a minority, like data-retrieval-only servers are non-mixed direct impact levels. We classify direct impact on tool level, with human validators agreement of 81\% (\emph{Fleiss'} \(\kappa\)\emph{=0.7)} on direct impact (perception vs. reasoning vs. action) and 85\% (\emph{Fleiss'} \(\kappa\)\emph{=0.5)} on functionality conditional on matching direct impact level. We classify direct impact on tool level, and then assign to a server the highest direct impact level of any of its tools. We ask human expert validators to directly assess direct impact on server level and find strong agreement (78\% agreement, \emph{Fleiss'} \(\kappa\)\emph{=0.5)}.

Human validators (n=6) agree 72\% (\emph{Fleiss'} \(\kappa\)\emph{=0.3)} with Sonnet 4.5's generality classification.

\subsection{Hierarchical classification methodology}
\label{app:hierarchy}

Adapted from \citet[supplementary materials]{handa2025which}.

A key challenge in mapping MCP tools to occupational tasks is the size of the O*NET task database. With 18,796 task descriptions across all occupations, direct classification via zero or few-shot prompting is impossible because the full list of tasks does not fit in the model\textquotesingle s context window. We instead construct this as a classification over a hierarchy of task labels, inspired by \citet{morin2005hierarchical} and \citet{mnih2008scalable}.

Our approach consists of three main components: (1) creating a hierarchical taxonomy of O*NET occupational tasks, (2) generating descriptive names for mid-level clusters, and (3) mapping MCP tools to O*NET tasks via hierarchical traversal.

\paragraph{(1) Creating a Task hierarchy}\label{creating-a-task-hierarchy}

We construct a three-level hierarchy with the base as the O*NET tasks using embedding-based clustering and semantic assignment:

\textbf{Step 1:} Embed task names. We embed all 18,796 O*NET task descriptions using the stella\_en\_400M\_v5 sentence transformer, obtaining 1024-dimensional vector representations for each task. Task text is augmented with occupation context by concatenating the task description with its associated occupation title (e.g., "Analyze financial data {[}Financial Analyst{]}").

\textbf{Step 2:} Create Level 2 clusters. We apply K-means clustering (k=400, n\_init=10) to the task embeddings, producing 400 mid-level clusters. Each cluster contains an average of 47 semantically related tasks. Cluster sizes range from 12 to 142 tasks.

\textbf{Step 3:} Assign Level 2 clusters to Level 1 categories. Unlike \citet{handa2025which}, who use iterative LLM-based hierarchy generation, we employ semantic cosine similarity assignment to 12 predefined high-level categories. Specifically:

\begin{itemize}
\item
 a) We embed the 12 Level 1 category names using the same sentence transformer.
\item
 b) For each Level 2 cluster, we compute the cosine similarity between its centroid and each Level 1 category embedding.
\item
 c) Each Level 2 cluster is assigned to the Level 1 category with highest similarity.
\end{itemize}

This approach yields an average assignment similarity of 0.61 (min: 0.45, max: 0.82), indicating strong semantic alignment between clusters and categories.

\textbf{Step 4:} Generate Level 2 cluster names. For each Level 2 cluster, we prompt Claude Sonnet 4 to generate a concise descriptive name (6-13 words) based on the tasks within the cluster. To improve disambiguation, we employ contrastive naming: the prompt includes both the tasks within the cluster and boundary tasks from neighboring clusters, helping the model focus on distinguishing characteristics.

\paragraph{The 12 Level 1 Categories}\label{the-12-level-1-categories}

Our hierarchy uses the following 12 high-level categories:

\begin{itemize}
\item
 L1\_01: Business management, finance, and customer service operations
\item
 L1\_02: Comprehensive healthcare services and medical specialties
\item
 L1\_03: Manage education, HR, and professional development programs
\item
 L1\_04: Design, implement, and maintain diverse information technology systems
\item
 L1\_05: Operate and manage diverse industrial and agricultural processes
\item
 L1\_06: Perform government regulatory enforcement and public safety operations
\item
 L1\_07: Conduct scientific research and technical analysis across disciplines
\item
 L1\_08: Create and preserve art, culture, and religious artifacts
\item
 L1\_09: Coordinate transportation networks and manage logistics supply chains
\item
 L1\_10: Manage diverse energy sources and optimize power systems
\item
 L1\_11: Design and construct infrastructure projects and engineering systems
\item
 L1\_12: Manage and improve environmental systems and sustainability practices
\end{itemize}

\paragraph{(2) Mapping MCP Tools to O*NET Tasks}\label{mapping-mcp-tools-to-onet-tasks}

To classify each MCP tool, we perform a tree-based search through the hierarchy:

\textbf{Step 1:} Level 1 selection. The LLM is presented with the MCP tool (name, description, input schema) and the 12 Level 1 categories, and selects the most appropriate category.

\textbf{Step 2:} Level 2 selection. Given the chosen Level 1 category, the LLM selects from among the Level 2 clusters assigned to that category (approximately 33 options on average).

\textbf{Step 3:} Task selection. Given the chosen Level 2 cluster, the LLM selects the single most appropriate O*NET task from within that cluster (approximately 47 options on average).

This hierarchical approach reduces the effective search space from 18,796 options to approximately 12 + 33 + 47 = 92 options across three tractable classification steps. Complete prompts are provided in Appendix~\ref{app:onet-prompt}.

\paragraph{Example Hierarchy Structure}\label{example-hierarchy-structure}

Figure A1 illustrates a subsection of the generated O*NET task hierarchy:

+-\/- L1\_01: Business management, finance, and customer service operations

\textbar{} +-\/- L2\_042: Financial analysis and investment portfolio management

\textbar{} \textbar{} +-\/- Analyze financial information to produce forecasts

\textbar{} \textbar{} +-\/- Evaluate investment performance and risk

\textbar{} \textbar{} +-\/- Prepare financial reports for stakeholders

\textbar{} \textbar{} +-\/- {[}44 additional tasks...{]}

\textbar{} +-\/- L2\_127: Customer service and client relationship management

\textbar{} \textbar{} +-\/- Respond to customer inquiries and complaints

\textbar{} \textbar{} +-\/- Process orders and handle transactions

\textbar{} \textbar{} +-\/- {[}38 additional tasks...{]}

\textbar{} +-\/- {[}31 additional L2 clusters...{]}

+-\/- L1\_04: Design, implement, and maintain diverse IT systems

+-\/- {[}10 additional L1 categories...{]}

\emph{Figure A1: Example subsection of the generated O*NET task hierarchy. Our hierarchy contains three levels: 12 top-level categories, 400 middle-level clusters, and 18,796 base-level O*NET tasks.}

Our approach differs from \citet{handa2025which} in three ways:

\textbf{1.} Level 1 assignment method: We use semantic cosine similarity to predefined categories rather than LLM-based iterative hierarchy generation. This provides deterministic, reproducible assignments while maintaining strong semantic coherence.

\textbf{2.} Embedding model: We use stella\_en\_400M\_v5 (1024 dimensions) rather than all-mpnet-base-v2 (768 dimensions), providing higher-dimensional representations.

\textbf{3.} Fixed hierarchy structure: Our 12 Level 1 categories are predefined rather than emergent from the clustering process, ensuring consistency with the Anthropic framework while simplifying the pipeline.

\paragraph{Connecting O*NET Tasks to Occupations}\label{connecting-onet-tasks-to-occupations}

The O*NET database covers 1,016 occupations across 23 major occupational groups (following the Standard Occupational Classification system). This occupational mapping enables analysis of which professions have the most MCP tool support, identification of occupational gaps in AI tooling, and comparison of tool availability across consequentiality of occupations. We aggregate tasks to occupations proportionally.

\subsection{AI-Created Server Detection}
\label{app:ai-detection}

We detect AI-created MCP servers through evidence in each repository's GitHub metadata. For each server, we query three sources via the GitHub REST API: the full commit history (up to 10,000 commits), up to 30 recent pull requests, and the recursive file tree. A server is classified as \texttt{ai\_authored = "yes"} if any of the following four criteria is met.

\paragraph{Criterion 1: Co-Authored-By Trailers.}
At least one \texttt{Co-Authored-By} trailer in any commit message or PR body references a known AI tool. AI coding agents add these trailers automatically, making them the most reliable indicator. Detected tools include Claude Code, GitHub Copilot, ChatGPT, Devin, Codex, Aider, Cline, Roo Code, Augment, Continue.dev, Gemini, and Windsurf.

\paragraph{Criterion 2: Configuration Files.}
At least one AI tool configuration file is present in the file tree, e.g.\ \texttt{CLAUDE.md}, \texttt{.cursor/}, \texttt{.github/copilot-instructions.md},
\texttt{.aider.conf.yml}, \texttt{.windsurfrules}, \texttt{.clinerules}, \texttt{.roo/}, \texttt{AGENTS.md}, \texttt{.augment/}, or \texttt{.continue/}.

\paragraph{Criterion 3: Bot Contributors.}
At least one commit or PR author matches a known AI bot account, such as \texttt{devin-ai-integration[bot]} or \texttt{copilot[bot]}. Dependency management bots
(dependabot, renovate) are excluded.

\paragraph{Criterion 4: AI Tool Mentions.}
At least one mention of an AI tool handle or name (e.g.\ \texttt{@claude}, \texttt{claude code}, \texttt{@copilot}) in commit messages or PR text. This criterion carries the highest false-positive risk due to ambiguous common words.

\paragraph{Agent Identification.}
We identify the most likely AI agent per server using weighted scores: configuration files (weight 10), bot contributors (5), Co-Authored-By matches (3), and mentions (1). The tool with the highest score is reported.

\paragraph{First-Month Analysis.}
To distinguish AI assistance present from the start of development from tools adopted later, we compute a first-month variant of the detection. For each server with a known \texttt{created\_at} date, we re-apply the same four criteria but restrict evidence to commits and pull requests dated within 30 days of repository creation. For configuration files, we check the file tree at the newest commit within this 30-day window rather than the latest tree, so that files added months later do not count. Servers without a \texttt{created\_at} date are excluded from first-month analysis. The agent identification scoring is applied separately to first-month evidence. Results in Section~\ref{sec:ai-created-results} use this first-month measure. We also record \texttt{date\_first\_ai\_evidence} as the earliest dated evidence across all four criteria: commit co-author trailers, AI handle mentions, bot contributors, and configuration file introduction dates (looked up via commit history).

\paragraph{Limitations.}
The approach cannot detect AI use that leaves no trace in git history, such as copying from a web chat interface. Squash merges may hide Co-Authored-By trailers. Configuration files may be added after initial development. We capture pre-October 2025 servers' READMEs in October 2025, and do not update it to understand original README and tooling. Newer servers are included with latest February 2026 READMEs. The 10,000-commit cap means the oldest commits in very large repositories go unchecked, though few MCP servers exceed this.

\paragraph{Cross-Validation with Pangram API.}
To assess to what extent we underestimate non-labelled AI assistance, we use the Pangram commercial AI content detection API on a subset of 197 repositories, analysing whether their READMEs are AI-created. Our conservative approach captures 28.2\% of repositories with READMEs classified by Pangram as significantly AI-generated (53.3\% binary agreement). Pangram analyses README text using
windowed classification, returning a \texttt{fraction\_ai} score (0.0--1.0). Our approach
analyses commit history, file trees, and contributor accounts. The two methods measure
different constructs: Pangram detects whether \emph{documentation text} was AI-generated,
while commit mining detects whether \emph{AI coding agents} were used in development.

Pangram classified 85 of 197 repositories (43.1\%) as AI-generated (\texttt{fraction\_ai} $\geq$ 0.8), with 76 (38.6\%) scoring above 0.9. The score distribution is bimodal: 41.5\% scored exactly 1.0 and 25.0\% scored exactly 0.0. Our commit-mining approach classified 55 of 197 (27.9\%) as AI-created, identifying Claude as the dominant tool (33 of 55, 60.0\%), followed by Copilot (10, 18.2\%) and Cursor (7, 12.7\%). The most common triggering criteria were AI handle mentions
(37 repositories), configuration files (27), and Co-Authored-By lines (26).

At the 0.8 threshold, binary agreement was 53.3\%: both methods agreed on 24 repositories as AI-involved and 81 as human-only. Our approach captured 24 of 85  Pangram-flagged repositories (28.2\% recall). The main source of disagreement was 61 repositories (31.0\%) where Pangram detected AI-generated README content but commit history showed no coding agent evidence, consistent with developers using conversational
AI (e.g.\ ChatGPT) for documentation without adopting agentic coding tools that leave commit-level traces. The reverse pattern (31 repositories, 15.7\%) suggests AI coding agents might be used but humans produce documentation.

\subsection{LLM Prompts}
\label{app:prompts}

\subsubsection{Processing readme to extract tools, MCP server validity and cleaned readme}

Filter README content and extract structured information useful for embedding analysis and consequentiality scoring.

Step 1: Create filtered\_content

KEEP in filtered\_content: Tool features and functionality, API docs and capabilities, Use cases and application areas, Integrations and connected services, Sector- or task-specific context

REMOVE from filtered\_content: Install/setup commands (e.g., npm, pip, docker), Prerequisites or system requirements, Code examples for setup/config, Directory layout, license, contributing, All URLs e.g. [GitHub](https://github.com) $\rightarrow$ GitHub

Step 2: CLASSIFY server type:

- is\_mcp\_server: 1 if this is an actual MCP server with tools/capabilities, 0 if it's just documentation, links, or references to MCP servers

Step 3:

EXTRACT tools information (try to copy the relevant exact text from the README):\\
Identify each distinct tool/function/capability mentioned,\\
Extract name and description for each tool,\\
Look for tool definitions, API endpoints, functions, commands, etc.

OUTPUT: Valid JSON object only, with this exact structure:

\begin{lstlisting}
{
  "summary": "Brief 1 sentence summary ...",
  "is_mcp_server": 1,
  "filtered_content": "Clean markdown content ...",
  "tools": [
    {
      "name": "first_tool_name",
      "description": "what this specific tool does"
    },
    {
      "name": "second_tool_name",
      "description": "what this other tool does"
    }
  ]
}
\end{lstlisting}

CRITICAL: Output ONLY the JSON object - no explanations, comments, or additional text.

GUIDELINES:

1. Preserve markdown format in filtered\_content

2. Focus on WHAT the tool does, not HOW to install

3. Set is\_mcp\_server to 1 for actual MCP servers, 0 for lists/references

4. Extract ALL distinct tools/functions/capabilities mentioned - create separate entries for each tool

5. Each tool should have a unique name - never duplicate tool names

6. If no specific tools are found, tools array can be empty

7. Tool descriptions should be copied, plus required input data such as API-keys

Original README content: [...]

\subsubsection{Prompt for MCP server classification on generality and payment autonomy}
\textit{Note: We collect data on generality across industry, and environment, but mainly use data on generality in the environment in this paper. }
You are a technology analyst identifying AI agent tools on model context protocol (MCP) servers. Follow this scoring process:

STEP 1: First, write analysis\_notes

Analyze the MCP server and document your findings in the analysis\_notes field. This should include:

- What the tool does and does not, any uncertainty about this

STEP 2: DESCRIBE action space:

- action\_space\_description: Describe what action space/environment this server operates in and its focus

* Example 1: "Pre-defined bitcoin wallet operations, in coinbase environment via API"

* Example 2: "General-purpose file system access across any directory and file type"

STEP 3: CLASSIFY generality (based on action space):

- generality\_industry: 1 if cross-industry (usable across many sectors), 0 if industry-specific

* Examples of 1 (cross-industry): desktop commander, file management, email client, calendar tools, note-taking

* Examples of 0 (industry-specific): crypto transaction tools, payment platforms, medical records systems, legal document processors

- generality\_environment: 1 if open/unconstrained environment, 0 if constrained/pre-specified environment

* Examples of 1 (open/unconstrained): computer use, browser automation, web scraping, generic file system access

* Examples of 0 (constrained): specific API (Stripe, GitHub, Slack), internal database, pre-configured service endpoint

STEP 4: Payment-specific analysis (only for servers with payment functionality):

- payments\_analysis: Text field describing which data is used to process payments from the MCP inputSchema or tool parameters. Examples: "amount, recipient\_address, payment\_method", "invoice\_id, card\_token", "wallet\_address, amount, gas\_price". Leave empty if no payment processing capability.

- payments\_autonomy: Rank the payment autonomy level (0-4):

* 0: Not a payment processing server (no payment functionality)

* 1: Only information about payments (invoice view, payment history, read-only)

* 2: Payment request or link created (generates payment links/requests but doesn't execute)

* 3: Payment processing via third-party (executes payments through external API like Stripe, PayPal)

* 4: Payment processing directly (direct payment execution with full control, e.g., blockchain transactions)

EXAMPLES:

Example 1 - Read-only banking monitor:

Input: asher-mcp

Description: Financial data aggregation tool

Tools:

- get\_accounts: Retrieve list of all connected bank accounts

- get\_account\_balance: Get current balance for a specific account

- get\_transactions: Retrieve transaction history for an account

- get\_investment\_holdings: View investment portfolio holdings

\begin{lstlisting}
{
  "server": "asher-mcp",
  "analysis_notes": "Read-only access to own bank accounts via scraping",
  "action_space_description": "Read-only access to connected bank accounts via financial data aggregation APIs",
  "generality_industry": 0,
  "generality_environment": 0,
  "payments_analysis": "not related to payments processing",
  "payments_autonomy": 1
}
\end{lstlisting}

Example 2 - Execution with limited transfer capabilities:

Input: base-mcp

Description: Blockchain interaction tool for Base network

Tools:

- get\_balance: Check wallet balance

- get\_transaction: Retrieve transaction details

- send\_transaction: Send ETH or tokens

- deploy\_contract: Deploy smart contracts

- interact\_contract: Call contract functions

- estimate\_gas: Calculate gas fees

Required inputs:

- private\_key: Wallet private key

- rpc\_endpoint: Base network RPC URL

\begin{lstlisting}
{
  "server": "base-mcp",
  "analysis_notes": "Readme is truncated, blockchain tool",
  "action_space_description": "Pre-defined blockchain operations on Base network via RPC endpoints",
  "generality_industry": 0,
  "generality_environment": 0,
  "payments_analysis": "wallet_address, private_key for signing already there, autonomous send_transaction tool without external approval",
  "payments_autonomy": 4
}
\end{lstlisting}

Example 3 - Poor documentation:

Input: ai-agent-mcp-servers

Description: Collection of MCP servers for AI agents

Tools: [No tool descriptions in documentation]

\begin{lstlisting}
{
  "server": "ai-agent-mcp-servers",
  "analysis_notes": "Almost no detail in the Readme",
  "is_finance_LLM": 0,
  "action_space_description": "Unclear - insufficient documentation to determine action space",
  "generality_industry": 1,
  "generality_environment": 1,
  "payments_analysis": "no data - assuming not for payments",
  "payments_autonomy": 0
}
\end{lstlisting}

Example 4 - General computer use:

Input: DesktopCommanderMCP

Description: Execute python and control mouse and keyboard on local OS

Tools: Tools:

- execute\_command: Execute arbitrary shell commands with timeout

- read\_file: Read file contents with pagination / negative offset

- write\_file: Write or append to files (line-limited)

- kill\_process: Terminate a running process by PID

\begin{lstlisting}
{
  "server": "DesktopCommanderMCP",
  "analysis_notes": "General-purpose MCP server for
    local automation: execute arbitrary terminal
    commands, manage processes, and perform full
    write operations on files. ...",
  "action_space_description": "General-purpose file
    system access across any directory and file type,
    with arbitrary command execution",
  "generality_industry": 1,
  "generality_environment": 1,
  "payments_analysis": "not payment focused",
  "payments_autonomy": 0
}
\end{lstlisting}

Output Format:

\begin{lstlisting}
{
  "server": "string",
  "analysis_notes": "Brief analysis of the tool(s)",
  "action_space_description": "Description of action
    space/environment ...",
  "generality_industry": 0|1,
  "generality_environment": 0|1,
  "payments_analysis": "string describing payment
    data fields used",
  "payments_autonomy": 0|1|2|3|4
}
\end{lstlisting}

\subsubsection{Prompt for hierarchical O*NET creation}
\label{app:onet-prompt}

Classify this MCP tool into an O*NET Level-1 cluster.

\texttt{<MCP tool name, description \& inputSchema>} 

For context, \texttt{<MCP Server name \& Description \& readme summary>}

Level 1 clusters:

\texttt{<id i: name i>}

\textit{see clusters in \ref{app:hierarchy}}

Respond with ONLY the id

\textit{Follow-up prompt:}

Level-1 chosen: \{l1\_id\}: \{l1\_name\}

Pick the SINGLE best Level-2 cluster id

\ldots{}

Level-2 chosen: \{l2\_id\}: \{l2\_name\}

Pick the SINGLE best Level-3 cluster id

\ldots{}

\subsubsection{Prompt for direct impact classification}

\texttt{<MCP tool name, description \& inputSchema>} For context, \texttt{<MCP Server name \& Description \& readme summary>}

Classify this MCP server tool by its direct impact and functionality

1. PERCEPTION (gathering information)

1.1 Sensors - database queries, monitoring, diagnostics, GUI reading, voice, search, physical sensing

2. REASONING (processing/analysis)

2.1 Planning - task decomposition, path-finding, workflow orchestration

2.2 Analysis - calculations, simulations, data processing

2.3 Resource Management - memory, self-management, resource allocation

3. ACTION (directly affecting the environment)

3.1 Authentication - login, CAPTCHA, wallet operations

3.2 Computer Use - GUI interaction, website automation, computer control

3.3 Code Execution - interpreters, IDE, file operations, running code

3.4 Software Extensions - calendar, social media APIs, third-party services

3.5 Physical Extensions - robotics, laboratory tools, physical world

3.6 Human Interaction - phone calls, messaging, direct communication

3.7 Agent Interaction - multi-agent coordination, sub-agents, third-party agents

Examples:

``get\_database\_records'' $\rightarrow$ 1.1

``calculate\_statistics'' $\rightarrow$ 2.2

``execute\_trade'' $\rightarrow$ 3.4

``run\_python\_code'' $\rightarrow$ 3.3

REPLY WITH NUMBER ONLY (e.g., 2.1) or `None' if unclear.

\subsection{Methods for monitoring agent use}
\label{app:monitoring}

\begin{table*}[htbp]
\centering
\caption{MCP monitoring is an early scanning method for monitoring of agentic deployments.}
\label{tab:monitoring}
\small
\resizebox{\textwidth}{!}{%
\begin{tabular}{>{\raggedright\arraybackslash}p{2.2cm}>{\raggedright\arraybackslash}p{3.2cm}>{\raggedright\arraybackslash}p{3.2cm}>{\raggedright\arraybackslash}p{3.2cm}>{\raggedright\arraybackslash}p{3.2cm}}
\toprule
\textbf{Criterion} & \textbf{Public agent tools}\newline(e.g. MCP on GitHub) & \textbf{Web scraping}\newline(news, jobs, websites) & \textbf{Interviews \& surveys}\newline(regulated companies) & \textbf{Agent usage data}\newline(model/agent providers) \\
\midrule
Early indicator & \textbf{Yes} E.g., GitHub release of Coinbase MCP wallet in January 2025, official in April 2025 & \textbf{Yes} E.g. Website announcement of Coinbase Agent wallet in Mid-2025 & \textbf{No} Months/Year later (E.g. Agent wallets not yet captured) & \textbf{Partly} Once used (E.g. API logs might show Coinbase Agent wallet usage) \\
Wide coverage & \textbf{Partly} Tools and their downloads, esp. for developers & \textbf{Partly} Agents and sometimes tools & \textbf{Partly} Agent and tools usage, specific to critical firms & \textbf{Yes} Depends on data agreements -- \citet{handa2025which} analysed API and claude.ai traffic \\
Precise coverage & \textbf{No} Usage statistics limited to non-caching downloads, difficult to identify specific user groups & \textbf{Partly} Website and news are linked to specific companies, may lack details & \textbf{Yes} Qualitative surveys can focus on user groups & \textbf{Yes} Accounts can be linked to user groups \\
Region-specific & \textbf{Partly} & \textbf{Partly} & \textbf{Yes} & \textbf{Yes} \\
High-stakes coverage & \textbf{Partly} Limited to public tools & \textbf{Partly} Limited to public announcements & \textbf{Yes} Might include private information in mandated qualitative surveys & \textbf{Partly} Highest-stakes agent uses likely not monitored by provider \\
Efficient \& available & \textbf{Yes} Can be automated, public, standardised format & \textbf{Partly} Some existing databases and providers, specific patterns require large-scale scraping & \textbf{No} Resource-intensive sourcing & \textbf{Partly} Can be automated, if access to private data is available \\
\midrule
\emph{Purpose} & \emph{\textbf{Early scanning}} & \emph{\textbf{Early scanning}} & \emph{\textbf{In-depth analysis}} & \emph{\textbf{Scanning \& In-depth analysis}} \\
\bottomrule
\end{tabular}%
}
\end{table*}

\subsection{Top-down usage domains}
\label{app:topdown-domains}

{\small
\setlength{\tabcolsep}{3pt}
\begin{longtable}{>{\raggedright\arraybackslash}p{1.6cm}>{\raggedright\arraybackslash}p{4.5cm}>{\raggedright\arraybackslash}p{2.5cm}>{\raggedright\arraybackslash}p{2.5cm}>{\raggedright\arraybackslash}p{1.8cm}>{\raggedright\arraybackslash}p{1.1cm}}
\caption{Task and occupation domains of AI agents in deployment}
\label{tab:topdown-domains} \\
\toprule
\textbf{Type} & \textbf{Cluster Label} & \textbf{Servers $n$}\newline\textbf{(\%, downloads \%)} & \textbf{Tools $n$}\newline\textbf{(\%, downloads \%)} & \textbf{Smithery}\newline\textbf{(tool use)} & \textbf{Claude.ai}\newline\textbf{usage} \\
\midrule
\endfirsthead
\toprule
\textbf{Type} & \textbf{Cluster Label} & \textbf{Servers $n$}\newline\textbf{(\%, downloads \%)} & \textbf{Tools $n$}\newline\textbf{(\%, downloads \%)} & \textbf{Smithery}\newline\textbf{(tool use)} & \textbf{Claude.ai}\newline\textbf{usage} \\
\midrule
\endhead
\midrule
\multicolumn{6}{r}{\textit{Continued on next page}} \\
\endfoot
\bottomrule
\endlastfoot
\rowcolor{rowgraydark}
Task domain & Design, implement, and maintain diverse information technology systems & 12,004 (68\%, 90\%) & 119,685 (67\%, 94\%) & 65\% (75\%) & 53\% \\
\rowcolor{white}
Occupation cluster (SOC) & Computer and mathematical occupations & 11,652 (66\%, 90\%) & 116,053 (65\%, 92\%) & 66\% (72\%) & 44.0\% \\
\rowcolor{rowgraydark}
Task domain & Create and preserve art, culture, and religious artifacts & 723 (4\%, $<$1\%) & 6,053 (3\%, $<$1\%) & 3\% (3\%) & 15\% \\
\rowcolor{white}
Occupation cluster (SOC) & Arts, design, entertainment, sports, and media occupations & 772 (4\%, 1\%) & 6,552 (4\%, $<$1\%) & 2\% (3\%) & 9.6\% \\
\rowcolor{rowgraydark}
Task domain & Business management, finance, and customer service operations & 2,397 (14\%, 5\%) & 31,882 (18\%, 4\%) & 7\% (15\%) & 11\% \\
\rowcolor{white}
Occupation cluster (SOC) & Office and administrative support occupations & 479 (3\%, 2\%) & 5,301 (3\%, 1\%) & 1\% (2\%) & 8.3\% \\
\rowcolor{rowgray}
Occupation cluster (SOC) & Business and financial operations occupations & 640 (4\%, 1\%) & 8,602 (5\%, 1\%) & 3\% (5\%) & 2.4\% \\
\rowcolor{white}
Occupation cluster (SOC) & Sales and related occupations & 1,290 (7\%, 2\%) & 17,126 (10\%, 2\%) & 3\% (7\%) & 2.1\% \\
\rowcolor{rowgray}
Occupation cluster (SOC) & Management occupations & 600 (3\%, 1\%) & 6,688 (4\%, 2\%) & 1\% (4\%) & 1.7\% \\
\rowcolor{white}
Occupation cluster (SOC) & Food preparation and serving related occupations & 107 (1\%, $<$1\%) & 1,174 (1\%, $<$1\%) & $<$1\% (0\%) & 0.4\% \\
\rowcolor{rowgraydark}
Task domain & Manage education, HR, and professional development programs & 201 (1\%, $<$1\%) & 1,857 (1\%, $<$1\%) & $<$1\% ($<$1\%) & 8\% \\
\rowcolor{white}
Occupation cluster (SOC) & Educational instruction and library occupations & 370 (2\%, $<$1\%) & 2,611 (1\%, $<$1\%) & 2\% (1\%) & 14.3\% \\
\rowcolor{rowgraydark}
Task domain & Conduct scientific research and technical analysis across disciplines & 1,273 (7\%, 3\%) & 8,989 (5\%, 1\%) & 23\% (6\%) & 6\% \\
\rowcolor{white}
Occupation cluster (SOC) & Life, physical, and social science occupations & 880 (5\%, 3\%) & 5,923 (3\%, 1\%) & 5\% (4\%) & 5.1\% \\
\rowcolor{rowgraydark}
Task domain & Perform government regulatory enforcement and public safety operations & 336 (2\%, $<$1\%) & 3,423 (2\%, $<$1\%) & $<$1\% ($<$1\%) & 2\% \\
\rowcolor{white}
Occupation cluster (SOC) & Legal occupations & 31 ($<$1\%, $<$1\%) & 294 ($<$1\%, $<$1\%) & 0\% (0\%) & 1.1\% \\
\rowcolor{rowgray}
Occupation cluster (SOC) & Protective service occupations & 141 (1\%, $<$1\%) & 1,676 (1\%, $<$1\%) & $<$1\% ($<$1\%) & 0.3\% \\
\rowcolor{rowgraydark}
Task domain & Operate and manage diverse industrial and agricultural processes & 59 ($<$1\%, $<$1\%) & 664 ($<$1\%, $<$1\%) & 0\% (0\%) & 2\% \\
\rowcolor{white}
Occupation cluster (SOC) & Production occupations & 79 ($<$1\%, $<$1\%) & 686 ($<$1\%, $<$1\%) & $<$1\% ($<$1\%) & 1.8\% \\
\rowcolor{rowgray}
Occupation cluster (SOC) & Installation, maintenance, and repair occupations & 8 ($<$1\%, $<$1\%) & 268 ($<$1\%, 0\%) & 0\% (0\%) & 0.5\% \\
\rowcolor{rowgraydark}
Task domain & Manage diverse energy sources and optimize power systems & 19 ($<$1\%, $<$1\%) & 199 ($<$1\%, $<$1\%) & 0\% (0\%) & 1\% \\
\rowcolor{rowgraydark}
Task domain & Manage and improve environmental systems and sustainability practices & 27 ($<$1\%, $<$1\%) & 232 ($<$1\%, $<$1\%) & $<$1\% ($<$1\%) & 1\% \\
\rowcolor{white}
Occupation cluster (SOC) & Building and grounds cleaning and maintenance occupations & 1 (0\%, 0\%) & 11 (0\%, 0\%) & 0\% (0\%) & 0.2\% \\
\rowcolor{rowgraydark}
Task domain & Comprehensive healthcare services and medical specialties & 186 (1\%, $<$1\%) & 1,642 (1\%, $<$1\%) & $<$1\% ($<$1\%) & 1\% \\
\rowcolor{white}
Occupation cluster (SOC) & Healthcare practitioners and technical occupations & 168 (1\%, $<$1\%) & 1,604 (1\%, $<$1\%) & $<$1\% ($<$1\%) & 1.7\% \\
\rowcolor{rowgray}
Occupation cluster (SOC) & Healthcare support occupations & 13 ($<$1\%, 0\%) & 72 ($<$1\%, 0\%) & 0\% (0\%) & 0.4\% \\
\rowcolor{white}
Occupation cluster (SOC) & Personal care and service occupations & 6 ($<$1\%, 0\%) & 107 ($<$1\%, 0\%) & 0\% (0\%) & 0.6\% \\
\rowcolor{rowgray}
Occupation cluster (SOC) & Community and social service occupations & 82 ($<$1\%, $<$1\%) & 532 ($<$1\%, $<$1\%) & $<$1\% ($<$1\%) & 2.4\% \\
\rowcolor{rowgraydark}
Task domain & Coordinate transportation networks and manage logistics supply chains & 190 (1\%, $<$1\%) & 1,358 (1\%, $<$1\%) & $<$1\% ($<$1\%) &  \\
\rowcolor{white}
Occupation cluster (SOC) & Transportation and material moving occupations & 57 ($<$1\%, $<$1\%) & 421 ($<$1\%, $<$1\%) & $<$1\% (0\%) & 0.4\% \\
\rowcolor{rowgraydark}
Task domain & Design and construct infrastructure projects and engineering systems & 114 (1\%, $<$1\%) & 1,450 (1\%, $<$1\%) & 0\% ($<$1\%) &  \\
\rowcolor{white}
Occupation cluster (SOC) & Architecture and engineering occupations & 143 (1\%, $<$1\%) & 1,627 (1\%, $<$1\%) & 16\% (1\%) & 1.5\% \\
\rowcolor{rowgray}
Occupation cluster (SOC) & Construction and extraction occupations & 8 ($<$1\%, 0\%) & 89 ($<$1\%, 0\%) & 0\% (0\%) & 0.2\% \\
\rowcolor{rowgraydark}
Task domain & Total Task domain & 17,529 (100\%, 100\%) & 177,434 (100\%, 100\%) & 100\% (100\%) & 100\% \\
\rowcolor{white}
Occupation cluster (SOC) & Total Occupation cluster (SOC) & 17,527 (100\%, 100\%) & 177,417 (100\%, 100\%) & 100\% (100\%) & 100.0\% \\
\end{longtable}
\vspace{1mm}
\begin{minipage}{\linewidth}
\scriptsize
\textit{Notes:} High-level task domains and below Standardised occupation classification (SOC) clusters as two approaches for domain mapping. See Section~\ref{sec:topic-modelling} for the two different methodologies. Usage is on server level, and assigned to tool level assuming 1 server use = 1 use of every tool on the server.
\end{minipage}
}

\subsection{Cumulative usage distribution}
\label{app:cumulative}

Usage is concentrated. For NPM downloads, the top 1\% (13 servers) cover 79.3\% of downloads, the top 10\% of servers cover 93.1\%. For PyPI downloads, the top 1\% (13 servers) dominate with 42.9\%, the top 10\% cover 74.5\%. We also considered using a `Use Count' statistic from Smithery's platforms but omitted it due to missing monthly data splits. Thus, our results are partly susceptible to misclassification of a few high-usage servers.

\begin{figure}[htbp]
\centering
\includegraphics[width=0.7\textwidth]{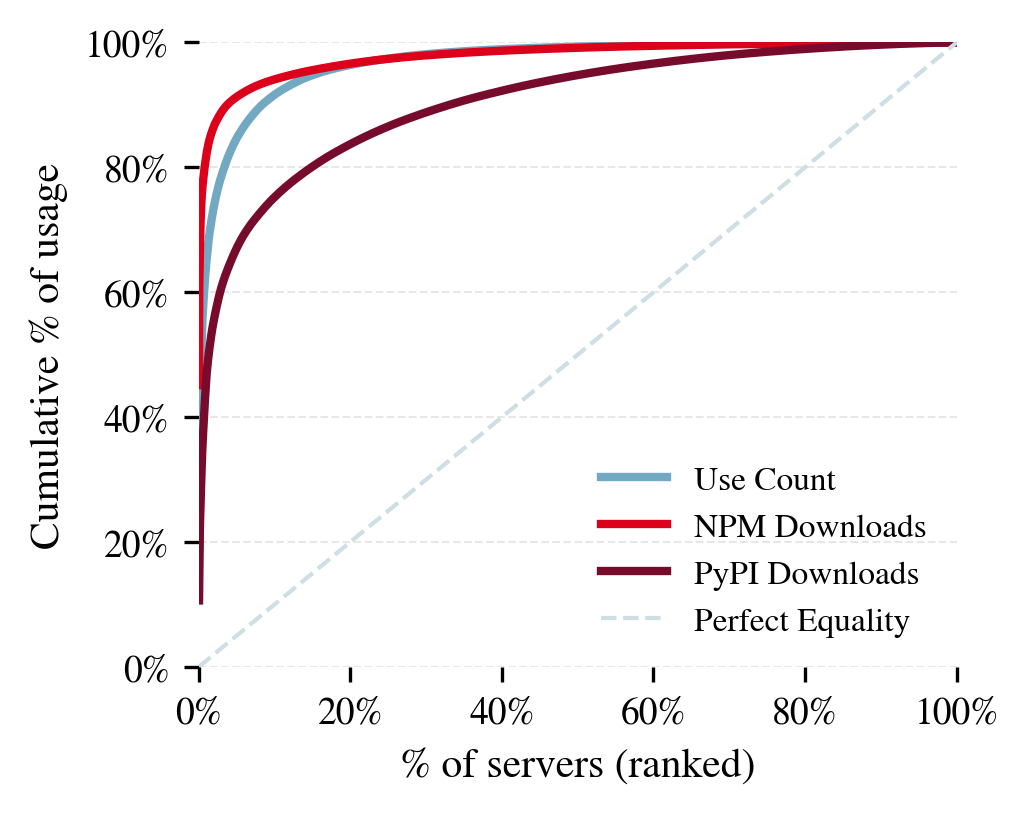}
\caption{\textbf{Cumulative usage distribution}, across total usage metrics (NPM, PyPI and Smithery's `use count'), by ranked MCP servers (only Servers with usage data included).}
\label{fig:cumulative-usage}
\end{figure}

\subsection{Bottom-up clustering to identify sub-clusters}
\label{app:subclusters}

\begin{figure}[htbp]
\centering
\includegraphics[width=\textwidth]{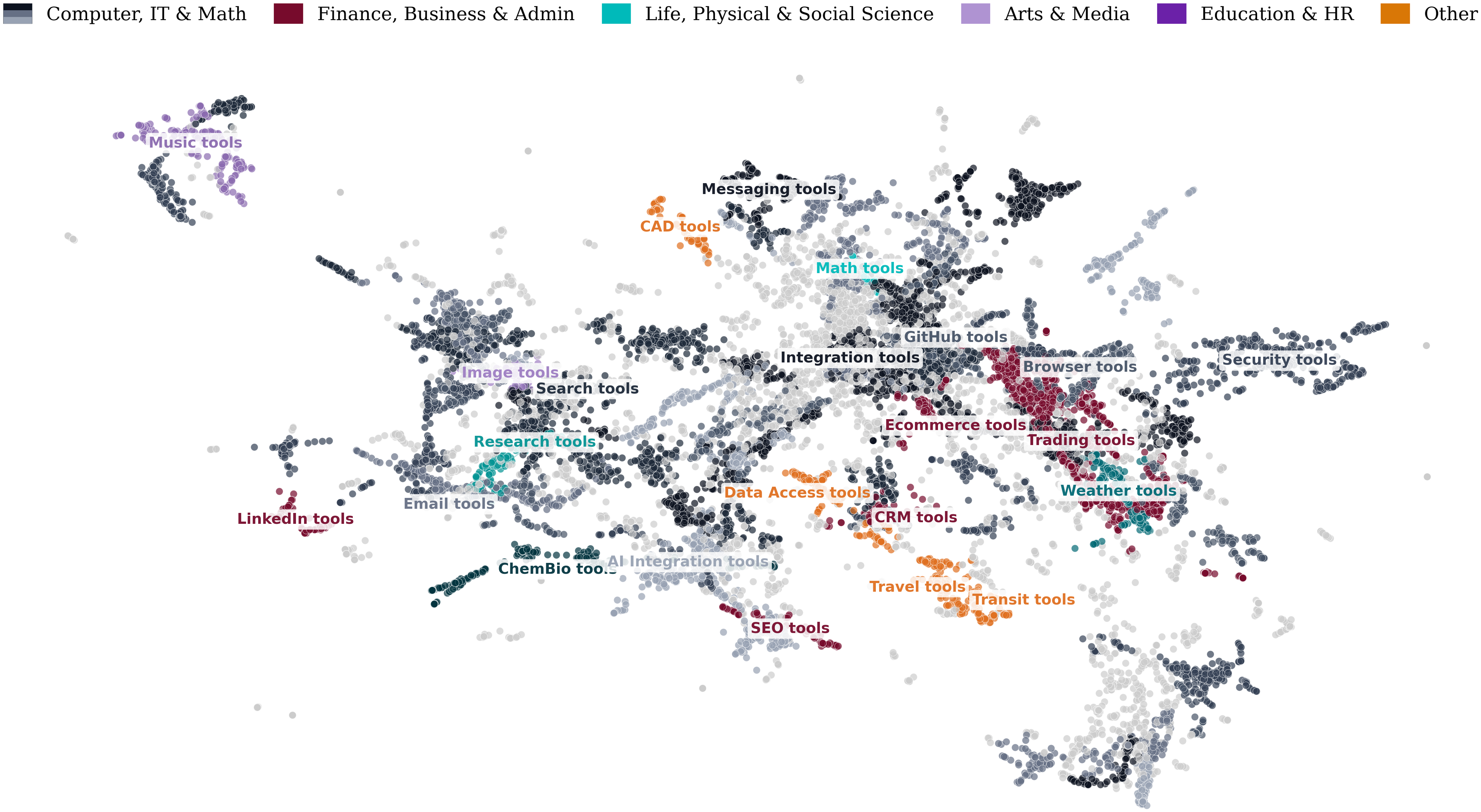}
\caption{\textbf{Bottom-up subclusters of MCP servers.} Each dot represents one MCP server. The legend and colouring show the match of top-down domains (see Section~\ref{sec:topic-modelling}) to the clusters. Labels provided for the top subclusters in each top-down domain. Clustering methodology in Section~\ref{sec:topic-modelling}.}
\label{fig:subclusters}
\end{figure}

\end{document}